\documentclass[twocolumn, tighten]{aastex631}

\def\teff{$T_{\rm eff}$}
\def\logg{$\log g$}
\def\bz{$\langle B_{\rm z}\rangle$}
\newcommand{\bs}{$\langle B_{\rm s}\rangle$}
\def\vs{$v_{\rm e}\sin i$}
\newcommand{\kms}{km\,s$^{-1}$}

\shorttitle{Empirical spectrum analysis of Nd~III}
\shortauthors{Ding et al.}
\graphicspath{{./}{figures/}}

\begin{document}

\title{The Spectrum and Energy Levels of the Low-lying Configurations of Nd~III}

\correspondingauthor{Milan Ding}
\email{milan.ding15@imperial.ac.uk}

\author[0000-0002-2469-8314]{Milan Ding}
\affiliation{Physics Department, Imperial College London, Prince Consort Road, London, SW7 2AZ, UK}

\author[0000-0003-2879-4140]{Juliet C. Pickering}
\affiliation{Physics Department, Imperial College London, Prince Consort Road, London, SW7 2AZ, UK}


\author[0000-0002-5321-5406]{Alexander N. Ryabtsev}
\affiliation{Institute of Spectroscopy, Russian Academy of Sciences, Troitsk, Moscow, 108840, Russia}

\author{Edward Y. Kononov}
\affiliation{Institute of Spectroscopy, Russian Academy of Sciences, Troitsk, Moscow, 108840, Russia}

\author[0000-0003-0551-1056]{Tatiana Ryabchikova}
\affiliation{Institute of Astronomy, Russian Academy of Sciences, Pyatnitskaya 48, Moscow, 119017, Russia}



\begin{abstract}
Emission spectra of neodymium (Nd, $Z=60$) were recorded using Penning and hollow cathode discharge lamps in the region 11500-54000 cm$^{-1}$ (8695-1852 \AA) by Fourier transform spectroscopy at resolving powers up to $10^6$. Wavenumber measurements were accurate to a few $10^{-3}$~cm$^{-1}$. Grating spectroscopy of Nd vacuum sliding sparks and stellar spectra were used to aid line and energy level identification. The classification of 433 transitions of doubly-ionised neodymium (Nd~III) from the Penning lamp spectra resulted in the determination of 144 energy levels of the $4\text{f}^4$, $4\text{f}^35\text{d}$, $4\text{f}^36\text{s}$, and $4\text{f}^36\text{p}$ configurations of Nd~III, 105 of which were experimentally established for the first time. Of the 40 previously published Nd~III levels, 1 was revised and 39 were confirmed. New Nd~III atomic structure calculations were made using the Cowan code parameterised by newly established levels. These results will not only benchmark and improve future semi-empirical atomic structure calculations of Nd~III, but also enable more reliable astrophysical applications of Nd~III, such as abundance analyses of kilonovae and chemically peculiar stars, and studies of pulsational wave propagation in these stars.
\end{abstract}


\section{Introduction} \label{sec:intro}
\begin{figure*} 
    \includegraphics[width=\linewidth]{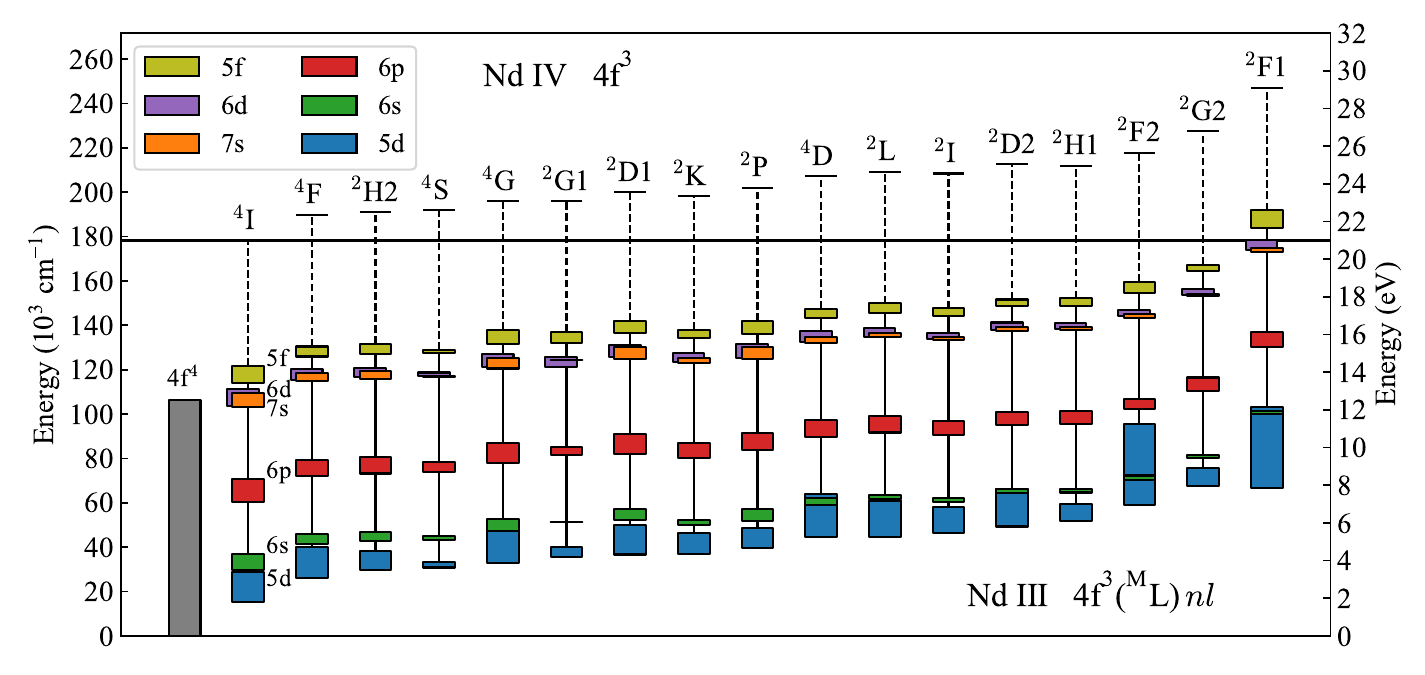}
    \label{fig1}
    \caption{Gross structure of the singly excited system of Nd~III, $4\text{f}^3(^{\text{M}}\text{L})nl$, up to $nl=\text{5f}$ and relative to the $LS$ terms of the Nd~IV ground configuration 4f\textsuperscript{3}. All levels are labeled according to the configurations and parent terms of their leading eigenvector components calculated during the present work using the Cowan code. The extent of each box merely shows the range of level energies within a given $4\text{f}^3(^{\text{M}}\text{L})nl$ sub-configuration. The ionisation potential is from \cite{johnson2017lanthanide} and the Nd~IV 4f$^3$ term energies are from \cite{wyart2007analysis}.}
\end{figure*}
The accuracy and availability of atomic data are crucial in the analyses of astrophysical spectra obtained by modern ground and space-based telescopes. The lanthanide elements ($Z=57-71$) are of great interest in the studies of nucleosynthesis, stellar evolution, and Galactic chemical evolution due to their still unsettled origins. Lines from the singly and doubly-ionised lanthanide species are present in spectra of hot chemically peculiar stars in large quantities, see for example, Przybylski's star \citep{przybylski1977iron, cowley2000abundances}. Numerous investigations on AT2017gfo \citep[e.g.,][]{valenti2017discovery, coulter2017swope}, the EM-counter part of neutron star merger event GW170817 \citep{abbott2017gw170817}, indicated large dependence of the kilonova opacity on lanthanide abundances \citep[e.g.,][]{kasen2013opacities,tanaka2013radiative, smartt2017kilonova, tanaka2018properties, watson2019identification}. However, the need for more extensive and accurate heavy element atomic data was almost always emphasised as the limitation to more detailed kilonova modeling. Much effort has been spent to address this by means of large-scale atomic structure calculations of the heavy elements \citep[e.g.,][]{kasen2013opacities, fontes2020line, tanaka2020systematic}. Nevertheless, reliable identification of kilonova spectral features is still hindered by the incompleteness of experimental atomic data and insufficient accuracies of theoretical calculations \citep{tanaka2020systematic, gillanders2021constraints}.

Despite the astrophysical interests in lanthanide spectral data, progress in obtaining accurate experimental energy levels and wavelengths of these elements is slow. Empirical investigations of lanthanide atomic structure are challenging due to their extremely complex and line-rich spectra owing to their $4\text{f}$ valence electrons. Amongst the first few doubly-ionised lanthanide elements, the spectrum of doubly-ionised neodymium (Nd~III, $Z=60$) is one of the most poorly known, with only 40 experimental energy levels established previously by \cite{ryabchikova2006rare}.

In this paper, we address the need for accurate lanthanide atomic data in astronomy, particularly for Nd~III, primarily by Fourier transform (FT) spectroscopy of a Penning discharge lamp \citep[e.g.,][]{finley1979continuous, heise1994radiometric} between 11500-54000~cm$^{-1}$ (8695-1852~\AA) using pure Nd cathodes. FT spectroscopy is capable of measuring spectra across a wide spectral range with high spectral resolution, ideal for investigating complex atomic spectra spanning from the IR to UV. At present, FT spectroscopy remains one of the best methods for extensive atomic energy level analysis. Recent examples include: 
Mn~II by \cite{liggins2021spectrum}, Ni~II by \cite{clear2022wavelengths}, and Zr~I-II by \cite{lawler2022energy}.

We present the empirical spectrum analysis for 144 energy levels of the $4\text{f}^4$, $4\text{f}^35\text{d}$, $4\text{f}^36\text{s}$, and $4\text{f}^36\text{p}$ configurations of Nd~III, 39 of 40 previously published energy levels were confirmed and 105 energy levels were established for the first time. In total, 433 Nd~III transitions were observed in the FT spectra with uncertainties down to a few $10^{-3}$~cm$^{-1}$. We also aim to provide insight into how accurate and extensive lanthanide atomic data was obtained using existing spectroscopic techniques. This research were substantially aided by FT spectroscopy of a Nd hollow cathode discharge, grating spectroscopy of Nd vacuum sliding sparks and Nd-rich stars, consideration of Nd energy level isotope shifts, and calculations using the Cowan code \citep{cowan1981theory,kramida2021suite} parameterised by newly established energy levels.

\section{Energy Level Structure of Nd III}
\subsection{Background}
Nd~III is of the neutral cerium isoelectronic sequence, with even-parity ground configuration [Xe]~$4\text{f}^4$ and singly excited configurations [Xe]~$4\text{f}^3nl$. Figure \ref{fig1} shows a schematic diagram of Nd~III energy levels of singly excited configurations up to $nl=\text{5f}$, grouped by parent terms of the 3-electron core under the LS-coupling scheme - $4\text{f}^3(^{\text{M}}\text{L})nl$. Transitions and levels involving the ground configuration and $nl =\:$5d, 6s, 6p are the focus of this paper.
The primary difficulty of investigating Nd~III atomic structure had been, and still remains to be, disentangling the immense number of spectral lines, particularly those of the 4f$^4$~-~4f$^3$5d transitions, as the 4f$^3$5d configuration is associated with hundreds of energy levels. In the visible-UV spectral region of the Nd discharge lamps, the 4f$^4$~-~4f$^3$5d lines of Nd~III also coincide with the complex spectra of Nd~I and Nd~II due to the low ionisation potentials of Nd. In most cases, empirical atomic structure investigations of the lanthanides require more extensive experimental methodology and data, and more labour intensive spectral analyses, compared to those of the lighter and less complex atoms with open subshells in the $l=\text{s},\text{p},\text{d}$ orbitals.

\subsection{Previous Experiments}
The first experimental investigations of Nd~III atomic structure were by \cite{dieke1961emission} and \cite{dieke1963spectra} using grating spectroscopy of spark discharges. This led to an unpublished establishment of 29 levels by H. M. Crosswhite, she had provided them for the \cite{martin1978atomic} compilation and they are found in the NIST database. The 29 levels account for classified transitions from 24 levels of $4\text{f}^35\text{d}$ to all 5 levels of the ground term $4\text{f}^4(^5\text{I})$, which were likely deduced from isolating Nd~III lines by comparing spectra of discharges at various temperatures \citep{dieke1961emission}. H. M. Crosswhite's unpublished lists of 314 classified Nd~III lines and 643 lines identified as belonging to Nd~III also circulate within the astrophysical community, and the latter proved useful for the present work. \cite{ryabchikova2006rare} later proposed revisions for 13 of the 29 levels and classifications for 11 brand new levels of the $4\text{f}^35\text{d}$ configuration, based on stellar spectroscopy, parameterised Cowan code calculations and Fourier transform spectroscopy of Nd hollow cathode discharges by \cite{aldenius2001}. 

\subsection{Energy Level Calculations}
Theoretical calculations of atomic structure and transition probabilities are invaluable in the experimental establishment of energy levels and classification of observed spectral lines. \cite{bord2000ab}, \cite{zhang2002measurement}, and \cite{dzuba2003energy} published the first theoretical investigations of the Nd~III spectrum, all utilising the Cowan code \citep{cowan1981theory}. Eigenvector compositions of the 4f$^3$5d levels below 33000~cm$^{-1}$ were published by \cite{zhang2002measurement} and \cite{ryabchikova2006rare} under the \textit{LS}-coupling scheme. \cite{gaigalas2019extended} and \cite{silva2022structure} published calculations for many more higher-lying levels and configurations, and the readily available online data from \cite{gaigalas2019extended} for level energies, leading \textit{LS} term labels, and transition probabilities proved useful during the initial stages of the present empirical spectrum analysis of Nd~III.

More accurate calculations of energy level values, level eigenvector compositions, and transition probabilities were required in the search for the experimentally unknown levels of Nd~III in the present work. The Cowan code \citep{cowan1981theory,kramida2021suite} was shown suitable for this purpose. The configuration interaction space consisted of the even parity configurations 4f$^4$ + 4f$^2$5d$^2$ + 4f$^3$6p + 4f$^3$5f and the odd parity configurations 4f$^3$5d + 4f$^3$6s + 4f$^3$7s + 4f$^3$6d + 4f$^2$5d6p. First, the calculations were performed in the \textit{ab initio} pseudo-relativistic Hartree-Fock (HFR) approximation with scaling factors of the Slater and configuration interaction parameters set at 0.85 and 0.70 respectively. To account for the incomplete interaction space, the Slater parameters and additional energy parameters $\alpha$, $\beta$, and $\gamma$ were then adjusted from considering calculations from \cite{ryabchikova2006rare} for Nd~III and from the calculations of the neighbouring spectra of Pr II \citep{mashonkina2009non}, Pr III \citep[][unpublished calculations]{palmeri2000theoretical,wyart2010abstract}, Nd II \citep[][unpublished calculations]{wyart2010theoretical,wyart2010abstract}, and Nd IV \citep{wyart2007analysis}. The energy parameters were adjusted in an iterative process by fitting newly established experimental energy levels when they became available during the empirical spectrum analysis.

The final energy parameters calculated for the 4f$^4$, 4f$^3$5d, 4f$^3$6s, and 4f$^3$6p configurations are listed in Table \ref{tab:cowan_params_4f}, remaining parameters are included in the machine-readable version online. The number of experimental levels was insufficient to fit all energy parameters; $\alpha$, $\beta$, and $\gamma$ were fixed at their estimated values for all configurations to allow the maximum number of Slater parameters to be fitted. There were enough core terms for all three F$^{2,4,6}$(4f,4f) electrostatic parameters only for the 4f$^3$5d configuration. It was possible to fit the levels of the 4f$^4$, 4f$^3$6p and 4f$^3$5f configurations with free F$^2$(4f,4f) and F$^4$(4f,4f) parameters and fixed F$^6$(4f,4f) parameter at the 0.852 ratio (Fitted/HFR) obtained for the 4f$^3$5d configuration. Only one core term ($^4$I) was available for the 4f$^3$6s, 4f$^3$7s, and 4f$^3$6d configurations, the F$^{2,4,6}$(4f,4f) of these configurations, and of the doubly-excited configurations 4f$^2$5d$^2$ and 4f$^2$5d6p, were fixed at their ratios obtained for the 4f$^3$5d configuration.
\startlongtable
\begin{deluxetable}{rlrcrr}
\tabletypesize{\scriptsize}
\tablecaption{Parameters of the least-squares fit of energy levels of the 4f$^4$, 4f$^3$5d, 4f$^3$6s, and 4f$^3$6p configurations of Nd~III of Cowan's codes. \label{tab:cowan_params_4f}}
\tablehead{
\colhead{Conf.} &
\colhead{Param.} &
\colhead{LSF\textsuperscript{a}} &
\colhead{G\textsuperscript{b}} &
\colhead{HFR\textsuperscript{a,c}} &
\colhead{Ratio\textsuperscript{a,d}}\\
\colhead{} &
\colhead{} &
\colhead{(cm$^{-1}$)} &
\colhead{} &
\colhead{(cm$^{-1}$)} &
\colhead{}
}
\startdata
4f$^4$  & $E_{\text{av}}$ & 32348(22)  &        & 35503        & -3156      \\
        & F$^2$(4f,4f)    & 71335(105) &        & 92636        & 0.770      \\
        & F$^4$(4f,4f)    & 40755(201) &        & 57680        & 0.707      \\
        & F$^6$(4f,4f)    & 35267(f)   & 3      & 41372        & 0.852      \\
        & $\alpha$(4f)    & 8(f)       &        & 0            &            \\
        & $\beta$(4f)     & -310(f)    &        & 0            &            \\
        & $\gamma$(4f)    & 1110(f)    &        & 0            &            \\
        & $\zeta$(4f)     & 782(4)     &        & 849          & 0.921      \\
\hline
4f$^3$5d& $E_{\text{av}}$ & 44999(43)  &        & 36770        & 8229       \\
        & F$^2$(4f,4f)    & 78070(397) & 1      & 101378       & 0.770      \\
        & F$^4$(4f,4f)    & 49616(383) & 2      & 63552        & 0.781      \\
        & F$^6$(4f,4f)    & 38949(554) & 3      & 45705        & 0.852      \\
        & $\alpha$(4f)    & 22(f)      &        & 0            &            \\
        & $\beta$(4f)     & -600(f)    &        & 0            &            \\
        & $\gamma$(4f)    & 1450(f)    &        & 0            &            \\
        & $\zeta$(4f)     & 882(2)     & 4      & 947          & 0.932      \\
        & $\zeta$(5d)     & 785(7)     &        & 831          & 0.945      \\
        & F$^1$(4f,5d)    & 953(81)    &        & 0            &            \\
        & F$^2$(4f,5d)    & 19310(119) &        & 26408        & 0.731      \\
        & F$^3$(4f,5d)    & 0(f)       &        & 0            &            \\
        & F$^4$(4f,5d)    & 13958(275) &        & 13022        & 1.072      \\
        & G$^1$(4f,5d)    & 9138(39)   &        & 12824        & 0.713      \\
        & G$^2$(4f,5d)    & 1689(198)  &        & 0            &            \\
        & G$^3$(4f,5d)    & 9205(142)  &        & 10202        & 0.902      \\
        & G$^4$(4f,5d)    & 0(f)       &        & 0            &            \\
        & G$^5$(4f,5d)    & 5148(126)  &        & 7735         & 0.666      \\
\hline
4f$^3$6s& $E_{\text{av}}$ & 55647(49)  &        & 47956        & 7691       \\
        & F$^2$(4f,4f)    & 78723(400) & 1      & 102226       & 0.770      \\
        & F$^4$(4f,4f)    & 50064(387) & 2      & 64126        & 0.781      \\
        & F$^6$(4f,4f)    & 39310(559) & 3      & 46129        & 0.852      \\
        & $\alpha$        & 22(f)      &        & 0            &            \\
        & $\beta$         & -600(f)    &        & 0            &            \\
        & $\gamma$        & 1450(f)    &        & 0            &            \\
        & $\zeta$(4f)     & 886(2)     & 4      & 953          & 0.932      \\
        & G$^3$(4f,6s)    & 2393(72)   &        & 2870         & 0.834      \\
\hline
4f$^3$6p& $E_{\text{av}}$ & 88343(29)  &        & 79811        & 8544       \\
        & F$^2$(4f,4f)    & 78655(146) &        & 102329       & 0.769      \\
        & F$^4$(4f,4f)    & 49923(164) &        & 64196        & 0.778      \\
        & F$^6$(4f,4f)    & 39344(f)   & 3      & 46180        & 0.852      \\
        & $\alpha$(4f)    & 22(f)      &        & 0            &            \\
        & $\beta$(4f)     & -600(f)    &        & 0            &            \\
        & $\gamma$(4f)    & 1450(f)    &        & 0            &            \\
        & $\zeta$(4f)     & 888(2)     &        & 954          & 0.930      \\
        & $\zeta$(6p)     & 2467(10)   &        & 1989         & 1.240      \\
        & F$^1$(4f,6p)    & 739(45)    &        & 0            &            \\
        & F$^2$(4f,6p)    & 6396(359)  &        & 7861         & 0.814      \\
        & G$^2$(4f,6p)    & 2082(55)   &        & 1981         & 1.051      \\
        & G$^3$(4f,6p)    & 0(f)       &        & 0            &            \\
        & G$^4$(4f,6p)    & 1490(93)   &        & 1791         & 0.832      
\enddata
\tablecomments{\\
\\\textsuperscript{a} Parameter values determined in the \textit{ab initio} pseudo-relativistic Hartree–Fock (HFR) and least-squares-fitted (LSF) calculations and their ratio. Standard deviation of the fitted LSF parameters are in brackets, where `f' means the parameter is fixed.
\\\textsuperscript{b} Group of the parameter, where parameters in each numbered group were linked together by sharing the same ratios to their corresponding HFR values.
\\\textsuperscript{c} The average energies were adjusted so that the energy of the ground level 4f$^4$ $^5$I$_4$ is zero in the HFR calculations with scaling of the Slater and configuration interaction parameters by factors of 0.85 and 0.70 respectively.
\\\textsuperscript{d} Differences between LSF and HFR parameters are given for $E_{\text{av}}$.
\\(This table is available in its entirety in machine-readable form)}
\end{deluxetable}
Level energies of the 4f$^2$5d$^2$ and 4f$^2$5d6p configurations overlap greatly with other configurations. Corresponding configuration interactions may affect calculated level energies and transition probabilities, especially in cases of close proximity to particular levels. Regrettably, no level was established in these doubly-excited configurations, and their energy parameters were adjusted only using information from neighbouring ion spectra and from the other Nd~III configurations. It should be noted that the estimated energies of the lowest $^5$L$_6$ levels in the 4f$^2$5d$^2$ and 4f$^2$5d6p configurations, 72827 and 126176~cm$^{-1}$ respectively, are in agreement with those predicted by \cite{brewer1971energies} at 72000$\pm$2000 and 126000$\pm$4000~cm$^{-1}$. 

Using the energy and configuration interaction parameters from Table \ref{tab:cowan_params_4f}, the standard deviations of the predicted level energies from their experimental values were 29 and 36 cm$^{-1}$ for the even and odd level systems respectively. Corresponding level eigenvector compositions and transition probabilities were also calculated. In searching for experimentally unknown energy levels, these calculations addressed the insufficient accuracy in the theoretical predictions by \cite{gaigalas2019extended}, particularly for higher-lying energy levels at higher level densities.

\section{Experimental Details} \label{sec:experiment}
\subsection{Fourier Transform Spectroscopy of Nd}
All level energies within this publication were determined using spectral lines measured by the high-resolution f/25 Imperial College VUV FT spectrometer \citep{thorne1987fourier}, using a custom-built water-cooled Penning discharge lamp (PDL), designed by \cite{heise1994radiometric}, with 99.5\% pure Nd cathodes (natural isotope abundances), a magnesium fluoride window, and argon as the sputtering carrier gas. The 0.5\% impurities in the cathodes were primarily iron and praseodymium, however, no known lines of these species were observed. Stable lamp running pressures and currents were chosen such that the signal-to-noise ratios (SNRs) of the previously classified known Nd~III transitions were maximised, which were (1.7-2.4)$\times 10^{-3}$~mbar and 650-750~mA respectively. The discharge remained stable for 2-3 hours under these conditions before the Nd cathodes were depleted. The PDL was chosen as the light source for its lower pressure and higher temperature stable discharge giving higher Nd~III populations, compared to those produced by the alternative hollow cathode discharge lamp commonly used in FT spectroscopy. As far as we know, no previous records on producing Nd spectra using a PDL light source have been published previously.

\begin{deluxetable*}{cccccccccr}
\tabletypesize{\small}
\tablecaption{Nd-Ar PDL FT spectra parameters.\label{tab:spec_params}}
\tablehead{
\colhead{Spec.} & \colhead{Spec. Range} & \colhead{Res.} & \colhead{Pressure} & \colhead{Current} & \colhead{No. of} & \colhead{Photomult.} & \colhead{Optical} & \colhead{Int. Calib.} & \colhead{$k$\textsubscript{eff}}\\[-6pt]
\colhead{} & \colhead{(cm$^{-1}$)} &  \colhead{(cm$^{-1}$)} & \colhead{($10^{-3}$~mbar)} & \colhead{(mA)} & \colhead{Co-adds} & \colhead{Tube} & \colhead{Filter} & \colhead{Lamp} & \colhead{($10^{-7}$)} \\
\colhead{(1)}&\colhead{(2)}&\colhead{(3)}&\colhead{(4)}&\colhead{(5)}&\colhead{(6)}&\colhead{(7)}&\colhead{(8)}&\colhead{(9)}&\colhead{(10)}
}
\startdata
A & 11890-17687 & 0.045 & 1.7 & 750     & 8  & R928   & OG530       & W & 1.74 $\pm$ 0.14\\
B & 17687-21300 & 0.050 & 2.2 & 750     & 7  & R11586 & GG475       & W & 1.39 $\pm$ 0.09\\
C & 21300-25369 & 0.050 & 2.3 & 650-750 & 19 & R11586 & GG385+SP500 & W & -3.10 $\pm$ 0.20\\
D & 25369-32480 & 0.055 & 2.4 & 750     & 21 & R11586 & UG5-C       & W & -2.95 $\pm$ 0.12\\
E & 32480-44422 & 0.070 & 1.7 & 750     & 23 & R7154  & -           & D$_2$ & -2.88 $\pm$ 0.19\\
F & 44422-53822 & 0.080 & 2.0 & 750     & 32 & R8486  & 180-B-1D    & D$_2$ & -4.32 $\pm$ 0.38\\
\enddata
\tablecomments{Column (1) is the spectrum name, column (2) is the spectral range within which the corresponding transitions were measured, column (3)-(5) are the spectral resolution, running pressures and currents respectively, column (6) is the number of interferogram co-adds, and columns (7)-(9) are the detector, filter, and intensity calibration lamp used respectively, W indicates tungsten lamp and D$_2$ indicates deuterium lamp. The final column lists the wavenumber calibration factors and their uncertainties.}
\end{deluxetable*}
Nd-Ar PDL FT spectra were recorded in six separate spectral regions within 11500-54000~cm$^{-1}$ (8695-1852~\AA). Spectral ranges spanning the visible were narrower (in~cm$^{-1}$) compared to the UV regions, this was to reveal weaker lines by reducing photon noise from the intense and numerous Nd~I-III lines in the visible. Spectral resolving power was limited by predominantly the Doppler line widths as no instrumental ringing was observed. Observed spectral lines from all 6 spectra were fitted to create a line list used in the energy level search. Most lines were fitted using the Voigt profile, where the Lorentzian components of the line widths were observed to be around one order of magnitude smaller than the Gaussian (Doppler) components. The statistical uncertainty in the wavenumber $\sigma_{\text{obs}}$ of a spectral line estimated by Voigt profile fitting is \citep{davis2001fourier}
\begin{equation}\label{eq:stat_unc}
    \Delta\sigma_{\text{obs}}=\frac{1}{\sqrt{N}}\frac{\text{FWHM}}{\text{SNR}},
\end{equation}
where $N$ is the number of points within the full width at half maximum $\text{FWHM}$, and $\text{SNR}$ was capped at 100 to prevent unrealistically low uncertainties. Details regarding non-Voigt profile spectral line fitting will be discussed in section \ref{subsec:wn_fit}.

Experimental details of the six Nd-Ar PDL FT spectra, labeled alphabetically, are listed in Table \ref{tab:spec_params}. 
The final column lists wavenumber calibration factors $k$\textsubscript{eff} of observed wavenumbers $\sigma$\textsubscript{obs} in each spectrum, where the calibrated wavenumbers $\sigma$\textsubscript{cal} are defined by \citep{haris2017critically}:
\begin{equation}\label{eq:wn_cal}
    \sigma\textsubscript{cal} = (1+k\textsubscript{eff})\sigma\textsubscript{obs}.
\end{equation}
This calibration factor mainly accounts for the finite aperture size of the FT spectrometer and any misalignment of the lamp and the spectrometer's calibration laser \citep{learner1988wavelength}. Wavenumber calibration was carried out using 13 of the 28 Ar~II reference lines recommended by \cite{learner1988wavelength} within spectrum C, using their corresponding reference wavenumbers measured by \cite{whaling1995argon}. The remaining Ar~II reference lines in spectrum C were unsuitable due to low SNRs and/or blending with Nd lines. 

A calibration factor $k_{\text{eff},i}$ was determined for each reference wavenumber $\sigma_{\text{cal},i}$ in a spectrum using its observed wavenumber in (\ref{eq:wn_cal}), and then the $k_{\text{eff}}$ of the spectrum was the weighted average
\begin{equation}
    k_{\text{eff}} = \frac{\sum_i w_i \, k_{\text{eff},i}}{\sum_i w_i},
\end{equation}
with weights $w_i=(\Delta k_{\text{eff}, i})^{-2}$, where the $\Delta k_{\text{eff}, i}$ were the absolute uncertainties of $k_{\text{eff},i}$ determined by (\ref{eq:wn_cal}). The uncertainty in the calibration factor $k_{\text{eff}}$ was given by \citep{radziemski1965arc, haris2017critically}
\begin{equation}\label{k_eff_unc}
    \Delta k_{\text{eff}} = \frac{\sqrt{\sum_i [w_i + w_i^2(k_{\text{eff}} - k_{\text{eff}, i})^2 ]}}{\sum_i w_i}.
\end{equation}
The systematic uncertainties $\Delta\sigma_{\text{sys}}$ in wavenumber calibration are given by $\sigma\textsubscript{obs}\Delta k\textsubscript{eff}$, which were added in quadrature with $\Delta\sigma_{\text{obs}}$ from (\ref{eq:stat_unc}) to yield the final wavenumber uncertainty for each line.

The above was repeated for each of the other spectra by using calibrated lines within overlapping spectral regions as wavenumber references. These were selected from lines that were non-blended, at intermediate SNRs, and with no evidence of self-absorption in their fit residuals. In this procedure, the $\Delta\sigma_{\text{sys}}$ of each other spectrum could become lower than that of spectrum C, e.g. low RMS residuals of $k_{\text{eff}, i}$, higher SNRs and/or a larger number of reference lines $i$. This was addressed by conservatively estimating the systematic uncertainty of a spectrum to be a linear sum including all previous systematic uncertainties, e.g.,
\begin{equation}\label{eq:sys_unc}            
\Delta\sigma_{\text{sys}}^{\text{F}} = \Delta\sigma_{\text{sys}}^{\text{F}} + \Delta\sigma_{\text{sys}}^{\text{E}} + \Delta\sigma_{\text{sys}}^{\text{D}} + \Delta\sigma_{\text{sys}}^{\text{C}},
\end{equation}
where the values from previous spectra were from overlapping spectral regions. This procedure was later verified by the lack of any systematic offsets seen between the observed and Ritz wavenumbers of transitions from energy levels with transitions across multiple spectra (Ritz wavenumbers are the differences between level energies optimised using all observed wavenumbers, see Section \ref{sec:energy_levels} for details). Differences between all observed and Ritz wavenumbers also lay within 1.2 times the estimated final wavenumber uncertainties, which were no more than 0.003~cm$^{-1}$ ($\sim$0.0003~\AA{} at 3000~\AA{}) for unblended lines with SNR $>$100 and free from self-absorption.

All six Nd-Ar PDL FT spectra were intensity calibrated using either a tungsten standard lamp (IR to UV) or a deuterium standard lamp (UV to VUV). Intensities measured in different spectral regions were placed onto a common relative intensity scale set by spectrum C, using lines within overlapping regions of neighbouring spectra. When possible, only known Nd~III lines were used. The observed relative line intensities from this work are recommended as only a rough guide, especially when comparing intensities between the different spectral regions listed in Table \ref{tab:spec_params}.

Spectra of a custom-built water-cooled hollow cathode discharge lamp (HCL) with the identical 99.5\% Nd cathode composition were also recorded using Ar carrier gas, with similar numbers of co-adds and the identical six spectral ranges, detectors, filters, and intensity calibration lamps listed in Table \ref{tab:spec_params}, but at higher resolutions due to the narrower Doppler line widths of the HCL spectral lines. Stable conditions were chosen to maximise the SNRs of previously classified Nd~III transitions at 0.43~mbar and 450~mA running pressure and current respectively. Under these conditions, the HCL ran stably for experiments over many weeks of use until cathode depletion. The HCL spectra were only used in this work as a guide for line intensities from a Nd-Ar plasma at a lower effective temperature compared to the Nd-Ar PDL plasma.

\subsection{Grating Spectroscopy of Nd}
Two sets of spectra of Nd vacuum sliding sparks (VS) of currents up to 1500~A were used for the analysis of Nd~III atomic structure. The spectra in the 390-1525~\AA{} and 1600-2536~\AA{} ranges were taken on the 6.65 m normal incidence spectrometer at the Institute of Spectroscopy in Troitsk. The spectrometer is equipped with a 1200~lines/mm grating providing a 1.25~\AA{}/mm plate factor. The spectra were recorded on Ilford Q photographic plates and measured on the automatic comparator with a scanning step of 0.3~µm controlled by the system for the automatic processing of photo-spectrograms designed by \cite{azarov1991system}. The other set of spectra consists of photographic plates from NIST, recorded on the 10~m normal incidence spectrograph and covered 2330-3250~\AA{}. These spectra were scanned by an Epson Expression 1000 XL scanner, and the spectral line positions and intensities were measured using the GFit code \citep{engstrom1998gfit}.

Impurity lines of oxygen, carbon, nitrogen, and silicon in various ionisation stages \citep{kramida2022nist}, as well as lines of Nd~IV \citep{wyart2007analysis} and Nd~V \citep{meftah2008spectrum}, were used for wavelength calibration initially. After the measurement and classification of Nd~III lines by FT spectroscopy in this work, the Nd VS grating spectra were calibrated again with the addition of the Nd III Ritz wavelengths. The final uncertainties of unblended and symmetric lines of moderate intensity were estimated at 0.006~\AA. 

In contrast to the HCL spectra, the Nd VS grating spectra were a guide for relative line intensities at higher effective temperatures compared to the PDL. Furthermore, these grating spectra also contained weaker Nd~III lines that were absent in the Nd-Ar PDL spectra.

\subsection{Stellar Spectroscopy of Nd}\label{sec:stellar spectra}
The laboratory analysis of the Nd~III spectrum was also supplemented by the analysis of high-resolution, high SNR spectra of the magnetic chemically peculiar Ap stars with large atmospheric Nd overabundance. Two tepid stars of similar atmospheric parameters (effective temperature and surface gravity) but different surface magnetic fields were chosen: HD~170973 with \teff=11200~K, \logg=3.8 \citep{2011mast.conf...69R} and HD~144897 with \teff=11250~K, \logg=4.0 \citep{ryabchikova2006rare}. The spectra of both stars ranged from 3030 to 10400~\AA{} and were recorded at resolving power R=80\,000 by the UVES instrument at the ESO VLT under the program 68.D-0254. More details of the observations and spectrum reduction are given in \cite{2008A&A...480..811R}. The surface magnetic field \bs=8.8~kG in HD~144897 was estimated from Zeeman splitting of spectral lines \citep{ryabchikova2006rare}, while the absence of any splitting in spectrum of HD~170973 together with one estimate of the longitudinal magnetic field 
\bz=392$\pm$5~G (G. Wade, private communication) did not indicate the presence of a surface magnetic field stronger than 1~kG in the atmosphere of HD~170973. Both stars rotate slowly with projected rotational velocity \vs=8.5~\kms\, (HD~170973) and \vs=3~\kms\, (HD~144897). 

The atmospheres of the two stars are very rich in rare-earth elements, their average Nd abundances\footnote{Nd abundance is expressed in a standard designation, $\log\varepsilon_{\rm Nd}$=$\log(N_{\rm Nd}/N_{\rm H})$ + 12, where $N_{\rm Nd}$ and $N_{\rm H}$ are number densities of neodymium and hydrogen respectively.} exceed the solar value by four orders of magnitude. In the solar photosphere, $\log\varepsilon_{\rm Nd}$=1.42 \citep{2021SSRv..217...44L}, whereas $\log\varepsilon_{\rm Nd}$=5.39$\pm$0.18 for HD~170973 \citep{2003PASJ...55.1133K}. In the present work, the reference Nd abundance of HD~170973 was recalculated at 5.63$\pm$0.14 using \teff=11200~K, \logg=3.8 \citep{2011mast.conf...69R}, and equivalent widths of 30 Nd~III lines previously classified by \cite{ryabchikova2006rare} and observed in the Nd-Ar PDL FT spectra. The reference abundance for HD~144897 was estimated at $\log\varepsilon_{\rm Nd}$=5.59$\pm$0.20 \citep{ryabchikova2006rare}. The temperatures, high Nd abundances, and differing surface magnetic field strengths of the two stars favoured the careful analysis of Nd~III lines in their spectra.
\begin{figure*}
    \includegraphics[width=\linewidth]{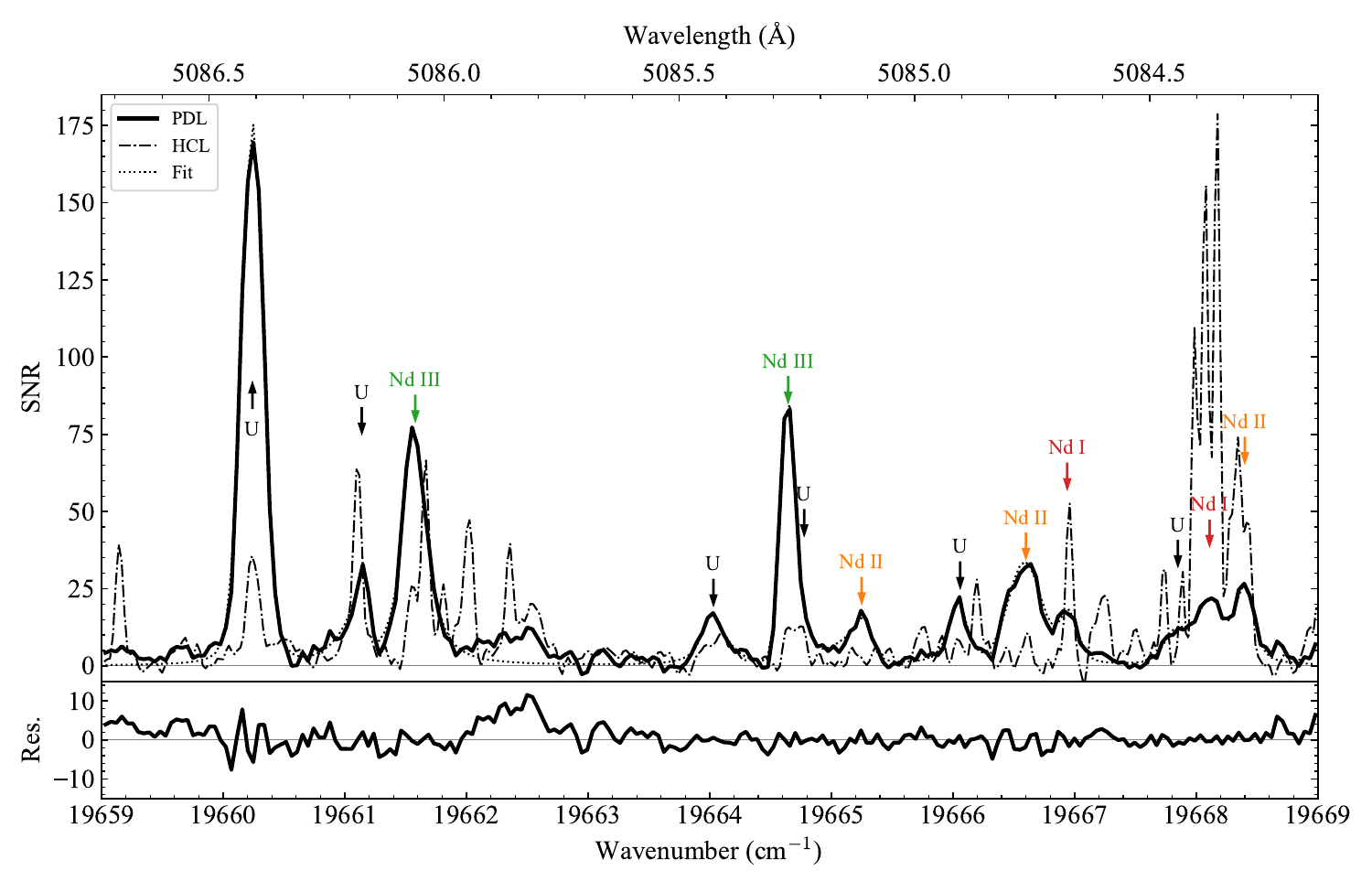}
    \caption{Emission spectrum of Nd recorded using Fourier transform spectroscopy between 19659-19669~cm$^{-1}$ (vacuum) containing lines with intermediate signal-to-noise ratios and showing weak and uncertain blended spectral lines. Penning discharge lamp (PDL) and hollow cathode lamp (HCL) spectra are solid and dash-dotted lines respectively. The dotted line shows fits to the 13 lines identified in the PDL spectrum, residuals are shown at the bottom. Lines identified and fitted in the PDL spectrum are indicated by vertical arrows and labels, those with label U are unclassified.} \label{fig2}
\end{figure*}

\section{Spectrum of the Nd Penning Discharge Lamp} \label{sec:spectrum}
\subsection{Spectrum and Line Identification}\label{sec:spec_line_id}
A total of 21584 lines were identified and fitted in the Nd-Ar PDL FT spectra by the peak-finding algorithm of \cite{nave2015xgremlin} and by manual fine-tuning. The line list contained transitions of Nd~I-IV and Ar~I-III and the numbers of observed known lines of each species are listed in Table \ref{tab:line_id}. Apart from Nd~III, the number of classified lines of each other species is uncertain due to the large quantity of unidentified blended and weak lines. For Nd~I and Nd~II, a significant number of matches with Ritz wavenumbers are expected to be coincidental and incorrect. Moreover, many levels from the NIST database are also uncertain with only configuration and $J$ value assignments. Figure \ref{fig2} highlights the difficulties and uncertainties of line identification with a 10~cm$^{-1}$ section of the Nd FT spectrum, measured in both the PDL and HCL and representative of the most line-rich and intense regions 15000-30000~cm$^{-1}$ (6666-3333~\AA), where a large fraction of the Nd~III 4f\textsuperscript{4}~-~4f\textsuperscript{3}5d, Nd~I, and Nd~II transitions lie. 
\begin{deluxetable}{lrc}
\tabletypesize{\footnotesize}
\tablecaption{Species and lines identified in Nd-Ar PDL FT spectra. \label{tab:line_id}}
\tablewidth{\linewidth}
\tablehead{
\colhead{Species} & \colhead{No. of Lines\textsuperscript{a}} & \colhead{Line List for Identification}
}
\startdata
Nd~I    & 1636 & NIST ASD\textsuperscript{b}  \\
Nd~II   & 6574 & NIST ASD\textsuperscript{b}  \\
Nd~III  & 580\textsuperscript{c}  & This work                    \\
Nd~IV   & 54   & \cite{wyart2007analysis}     \\
Ar~I    & 48   & \cite{whaling2002argon}      \\
Ar~II   & 430  & \cite{saloman2010energy}     \\
Ar~III  & 90   & \cite{saloman2010energy}     \\
\enddata
\tablecomments{\\ \textsuperscript{a} Number of lines whose wavenumbers matched known lines within $\pm 0.2$~cm$^{-1}$ for Nd~IV, and $\pm 0.05$~cm$^{-1}$ for the other species.
\\\textsuperscript{b} Using Ritz wavenumbers allowed by $\Delta J=0,\pm1$, but not $J=0\leftrightarrow0$ and parity selection rules for E1 transitions of the energy levels that were mostly from \cite{martin1978atomic,blaise1971present,blaise1984revised}.
\\\textsuperscript{c} 147 of these lines belong to the 4f$^3$6p~-~4f$^3$7s and 4f$^3$6p~-~4f$^3$6d transitions, which will be reported in a future publication.}
\end{deluxetable}

Weak lines and weak blends such as the feature at 19662.5~cm$^{-1}$ of the PDL spectrum in Figure \ref{fig2} were often missed by the peak finding algorithm and manual fine-tuning when fitting line profiles for the line list. The labour cost of analysing and fitting every single line around the noise level was considered not feasible due to their ubiquity and large uncertainties from high susceptibility to blending. For example, in the comparison between the PDL and HCL spectra of Figure \ref{fig2}, all 13 PDL lines of the 10~cm$^{-1}$ section were arguably blended to some extent. The incompleteness and inaccuracy of the list of observed wavenumbers were two of the most limiting factors in the empirical search for energy levels; expected transitions may be missing in the line list due to weak and unfitted lines or inaccurately fitted blends. Around 2000 of the 21559 lines were in fact fitted later during the empirical Nd~III energy level analysis when more thorough spectrum fitting of blends and weak lines became necessary within spectral regions containing weak Nd~III transitions of interest.

The primary assistance in line and blend identification was provided by the HCL spectra, which produced an Nd-Ar discharge at a lower temperature with narrower Doppler widths (this is more conveniently seen in the upper plot of Figure \ref{fig3}). Lines of Nd~I could also be identified with reasonable certainty, as Nd~I was the only Nd species with reduced relative intensities in the PDL compared to that of the HCL, and this is noticeable in Figure \ref{fig2}. Self-absorption was observed in the PDL for a fraction of Nd~II transitions between the lowest-lying levels, evident from the self-reversals, flattened tops, or most commonly the reversed `W' Voigt profile fit residuals of these line profiles. The line in Figure \ref{fig2} at 19660.2~cm~$^{-1}$ was unclassified but showed features of self-absorption, indicating the possibility of it being an unclassified Nd~II transition.

Compared to the line-rich visible-UV regions, the Nd-Ar PDL spectrum was relatively much less intense in the deeper UV (30000-50000~cm$^{-1}$, 3333-2000~\AA) and was not plagued by a large number of blended and weak lines. This indicated lower populations of Nd~III ions at the 4f$^3$6p, 4f$^3$7s, 4f$^3$6d and doubly excited configurations, and of Nd~IV ions at the 4f$^2$5d and 4f$^2$6p configurations; their transitions to lower levels were expected in this region but were observed at low SNRs. The lines with the highest SNRs in this region were the 4f$^4$~$^5$I - 4f$^3$($^4$G)5d transitions of Nd~III around 35500~cm$^{-1}$.

\subsection{Nd III Isotope Shifts}
Many Nd lines of the FT spectra showed similar non-Voigt line profiles despite the likelihood of line blending. The cathodes of the PDL and HCL contained natural abundances of the 7 stable isotopes \citep{berglund2011isotopic}: Nd-142 (27.15\%), Nd-143 (12.17\%), Nd-144 (23.80\%), Nd-145 (8.29\%), Nd-146 (17.19\%), Nd-148 (5.76\%), and Nd-150 (5.64\%), all of which have non-negligible abundance. Energy levels of Nd-143 and Nd-145 also have hyperfine structure. Consideration of isotope shifts of different transition arrays was a key component in the empirical atomic structure analyses of Nd~I and Nd~II \citep[e.g.,][]{blaise1971present, blaise1984revised, wyart2010theoretical}{}. Likewise, isotope line profiles of the 4f\textsuperscript{4}~-~4f\textsuperscript{3}5d Nd~III transitions were indispensable for their classifications in this work. 

\begin{figure}
    \includegraphics[width=\linewidth]{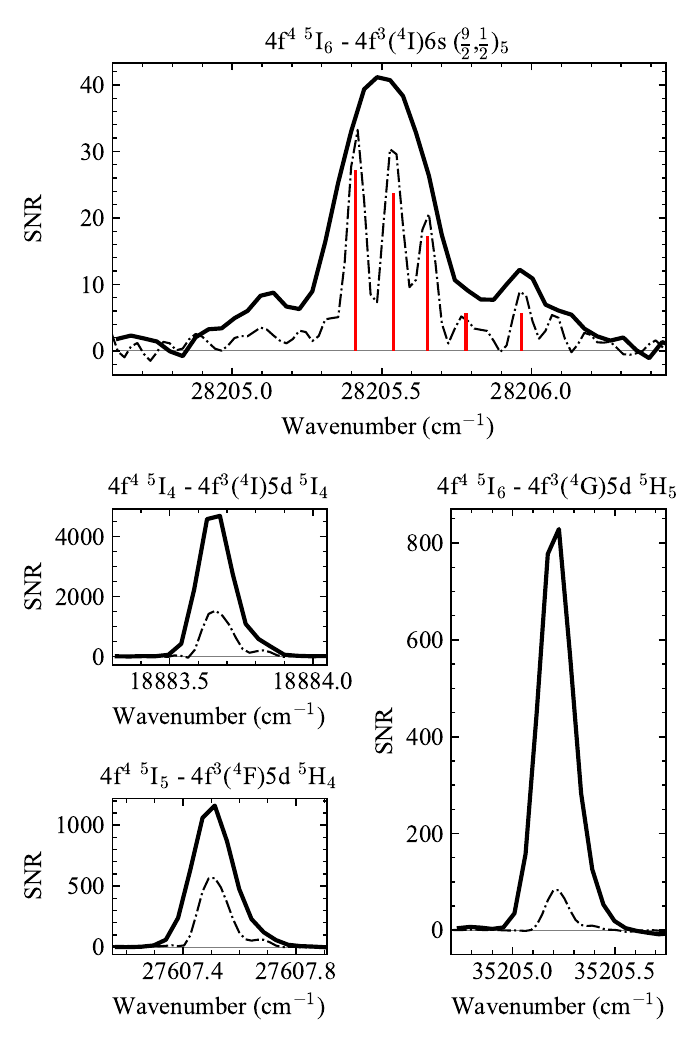}
    \caption{Isotope structure of 4 Nd~III lines observed by Fourier transform spectroscopy of the Nd-Ar Penning discharge lamp (PDL, solid) and the Nd-Ar hollow cathode lamp (HCL, dash-dotted). For 4f\textsuperscript{4}~\textsuperscript{5}I\textsubscript{6}~-~4f\textsuperscript{3}(\textsuperscript{4}I)6s~$(\frac{9}{2},\frac{1}{2})_5$, isotope components of Nd-142, Nd-144, Nd-146, Nd-148, and Nd-150 are marked tentatively from left to right with red vertical lines in relative abundance ratios from \cite{berglund2011isotopic}, remaining features are likely hyperfine structures of Nd-143 and Nd-145 or blended weak lines.} \label{fig3}
\end{figure}

Observations of Nd~III isotope structure were first made by \cite{aldenius2001}. Examples observed in the FT spectra of the present work are shown in Figure \ref{fig3}. The transition 4f\textsuperscript{4}~\textsuperscript{5}I\textsubscript{6}~-~4f\textsuperscript{3}(\textsuperscript{4}I)6s~$(\frac{9}{2},\frac{1}{2})_5$ was the only Nd~III line observed in the HCL spectrum with resolved Nd-142, Nd-144, and Nd-146 isotope components. The Nd-148 and Nd-150 components at higher wavenumbers are uncertain due to smaller relative abundances, low SNRs, and hyperfine structure of the Nd-143 and Nd-145 isotopes. This pattern was a reflection of, but consistent with, many isotope profiles observed for Nd~I (e.g., 19668.1~cm$^{-1}$ of Figure \ref{fig2}) and Nd~II \citep[e.g., Figure 2 of][]{koczorowski2005laser}{}{}. Isotope shifts added another layer of complexity in line identification; in Figure \ref{fig2}, the unclassified line of 19667.9~cm$^{-1}$ was possibly misidentified as an individual line rather than a part of the isotope structure of the Nd~I line. Almost all classified 4f\textsuperscript{4}~-~4f\textsuperscript{3}5d transitions of Nd~III showed an unresolved Nd-150 right-side wing within the PDL spectra (e.g., see Figure \ref{fig3}), which was an essential clue to their classification when observed at a sufficient SNR.

\subsection{Line Profile Fitting} \label{subsec:wn_fit}
Least-squares fitting using the Voigt profile was not suitable for all the observed PDL spectral line profiles. Lines showing significant isotope structure were fitted by using the centre-of-gravity (COG) wavenumber, i.e., an average of wavenumbers within a manually specified range weighted by SNR. For COG fitted lines, statistical uncertainties from equation (\ref{eq:stat_unc}) were doubled. However, the unresolved Nd-150 right-side wings of the 4f\textsuperscript{4}~-~4f\textsuperscript{3}5d Nd~III transitions were not always obvious during visual inspections and manual corrections to the output of the peak-finding algorithm, particularly at intermediate and lower SNRs. In this case, not only would the isotope shift clue be lost, but the differences between wavenumbers from Voigt and COG fits were often larger than their statistical uncertainties. This again caused inaccuracies of wavenumbers in the line list, which obstructed the search for Nd~III energy levels. Additionally, COG fitting may not be as accurate for blended and weak lines, as specifying wavenumber limits of such line profiles could range from non-trivial to impossible.

Instead, an asymmetric Voigt profile was used to quantify the unresolved Nd-150 right-side wings of the 4f\textsuperscript{4}~-~4f\textsuperscript{3}5d transitions, and to more accurately estimate wavenumbers of similarly asymmetric line profiles. The dominant Doppler width of the Voigt profile $w$ was varied as a sigmoid function of wavenumber parameterised by an asymmetry factor $a$ \citep{stancik2008simple},
\begin{equation}
    w(\sigma) = \frac{2w_0}{1 + e^{a (\sigma - \sigma_0)}},
\end{equation}
where $w_0$ and $\sigma_0$ are the Doppler width and wavenumber of the usual Voigt profile ($a=0$) respectively. The wavenumber of the fitted line was estimated using the COG of this asymmetric Voigt profile. Lines resembling the Nd-150 right-side wing asymmetry would have $a<0$.

\begin{figure}
    \includegraphics[width=\linewidth]{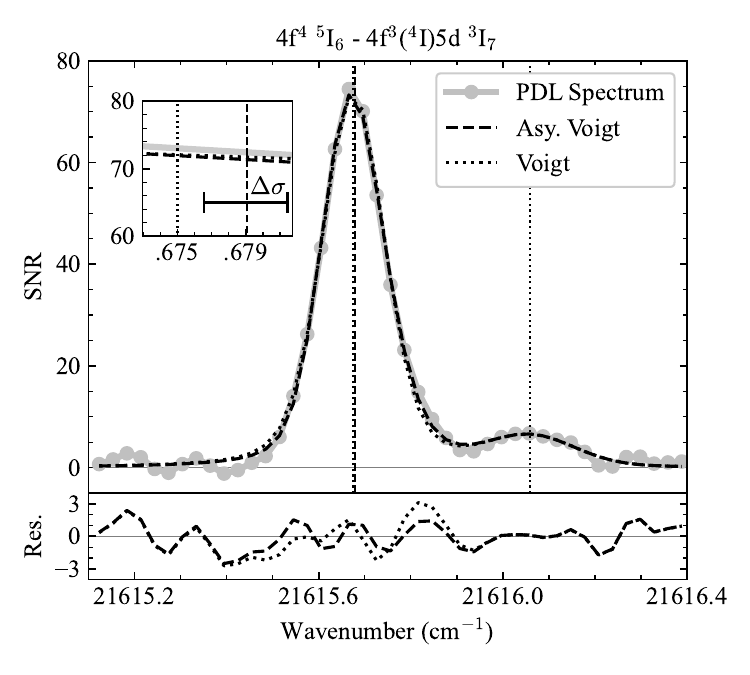}
    \caption{Nd~III transition 4f\textsuperscript{4}(\textsuperscript{5}I\textsubscript{6})~-~4f\textsuperscript{3}(\textsuperscript{4}I)5d~\textsuperscript{3}I\textsubscript{7} at 21615.679~cm$^{-1}$, slightly blended with an unclassified line at 21616.058~cm$^{-1}$. Fits using asymmetric Voigt and normal Voigt profiles are shown in dashed and dotted lines respectively, their residuals are shown in the lower plot and their fitted wavenumbers are shown by the vertical lines. The wavenumber uncertainty of the asymmetric Voigt fit $\Delta\sigma$ is shown in the zoomed section.} \label{fig4}
\end{figure}

By no means was this asymmetric Voigt profile an accurate physical description of isotope shifts, but overall improvements in wavenumber fitting were achieved for the 4f\textsuperscript{4}~-~4f\textsuperscript{3}5d transitions. In the example shown in Figure \ref{fig4}, the choice of COG wavenumber limits was difficult due to slight blending with a second weaker line, the unresolved right wing was not obvious from inspection and only evident from Voigt fit residuals. When fitted using the Voigt profile, the observed wavenumber differed from Ritz and asymmetric Voigt fit values by more than the wavenumber uncertainty. The example in Figure \ref{fig4} was fitted with $a = -1.62$. Statistical uncertainties from equation (\ref{eq:stat_unc}) were doubled for any lines with $|a|>1$, and $a$ was fixed at 0 for lines with SNRs lower than 10.

Compared to using the normal Voigt function, wavenumbers from asymmetric Voigt fits were closer to and within uncertainties of manual COG values. This fitting method also alleviated a prohibitive amount of labour otherwise induced by manually specifying limits of COG wavenumbers of thousands of such lines. Flexible adjustments of the line list using all 3 line fitting methods were readily available using a custom computer program, where each line (row) of the line list contained the essential parameters (columns): SNR, FWHM, calibrated wavenumber (Voigt or COG) and its final uncertainty, calibrated relative intensity, method of fitting, and the asymmetry factor $a$. 

\section{Empirical Spectrum Analysis of Nd III}\label{sec:term_analysis}
Initially, results from previously published experimental investigations of Nd~III were used to classify Nd~III lines, and then unclassified lines were used in the search for new energy levels using theoretical calculations as a guide. The methodology is discussed in more detail in Section \ref{sec:analysis_method}.

\subsection{Results}
\subsubsection{Energy Levels}\label{sec:energy_levels}
Wavenumbers of classified Nd~III transitions from the Nd-Ar PDL FT spectra were input into the program LOPT \citep{kramida2011program} to estimate the optimum level energies and Ritz wavenumbers by minimisation of the squared difference between observed wavenumbers and their Ritz values. These observed wavenumbers were calibrated onto the same absolute scale, but systematic wavenumber shifts (below the systematic uncertainties) still remain between each spectrum. Initially, the systematic shifts of wavenumbers from each spectrum were assumed zero for the input of LOPT, and systematic uncertainties from calibration were only included as a part of the weights in the best fit by LOPT, therefore risking underestimation of calibration uncertainties for the Ritz wavenumbers. To address this, the systematic uncertainties $\Delta \sigma_{\text{sys}}$ from the six spectra were added in quadrature (yielding $\pm0.0052$~cm$^{-1}$) and then in quadrature with all Ritz wavenumber uncertainties determined by LOPT for a more conservative estimate of line and level energy Ritz wavenumber uncertainties. 

The experimentally established energy levels of the 4f$^4$, 4f$^3$5d, 4f$^3$6s, and 4f$^3$6p configurations of Nd~III from this work are listed in Table \ref{tab:nd_ft_levels} in order of ascending energy of their assigned configurations and terms. 
Pure \textit{LS}-coupling basis functions were used to label energy levels of the 4f\textsuperscript{4}, 4f\textsuperscript{3}5d and 4f\textsuperscript{3}6p configurations. Under this scheme, purity is high overall for the lowest-lying terms of the 4f\textsuperscript{4} and 4\textsuperscript{3}5d configurations. Higher-lying levels of these two configurations, and the 4f\textsuperscript{3}6p levels, were mixed; violation of the \textit{LS}-coupling selection rules (i.e., $\Delta S = 0$ and $\Delta L = 0, \pm 1$ but not $L=0\leftrightarrow0$) was thus very common. For the 4f\textsuperscript{3}($^4$I)6s levels, \textit{jj}-coupling of the 6s electron with the 4f\textsuperscript{3}(\textsuperscript{4}I) core was used for assignment as the basis functions under this coupling scheme better represented their assigned levels.
\startlongtable
\begin{deluxetable*}{rcc|r|r|rr|c|rlrlrl}
\tabletypesize{\scriptsize}
\tablecaption{Energy levels experimentally established for the 4f$^4$, 4f$^3$5d, 4f$^3$6s, and 4f$^3$6p configurations of Nd~III.  \label{tab:nd_ft_levels}}
\tablehead{
\multicolumn{3}{c|}{Assignment} & 
\multicolumn{1}{c|}{Energy} &
\multicolumn{1}{c|}{$N$} &
\multicolumn{1}{c}{$\Delta E_1$} &
\multicolumn{1}{c|}{$\Delta E_2$} &
\multicolumn{1}{c|}{$g$}&
\multicolumn{6}{c}{Eigenvector Composition}
\\
\colhead{Conf.} & 
\multicolumn{1}{c}{Term} & 
\multicolumn{1}{c|}{$J$} & 
\multicolumn{1}{c|}{(cm$^{-1}$)}&
\multicolumn{1}{c|}{}&
\multicolumn{2}{c|}{(cm$^{-1}$)}& 
\multicolumn{1}{c|}{}&
\multicolumn{6}{c}{(\%)}
}
\startdata
4f$^4$ & $^5$I & 4 & 0.000(\;\:\,-) & 22(5) & -8 & 0 & 0.604 &  97 & 4f$^4$ $^5$I  &   1 & 4f$^4$ $^3$H4  &   1 & 4f$^4$ $^3$H3 \\
 &  & 5 & 1137.794(\;\:5) & 31(7) & -3 & 65 & 0.901 &  98 & 4f$^4$ $^5$I  &   1 & 4f$^4$ $^3$H4  &  &   \\
 &  & 6 & 2387.544(\;\:6) & 34(7) & -4 & 124 & 1.071 &  99 & 4f$^4$ $^5$I  &  &    &  &   \\
 &  & 7 & 3714.548(\;\:6) & 29(6) & -12 & 172 & 1.177 &  98 & 4f$^4$ $^5$I  &   1 & 4f$^4$ $^3$K2  &  &   \\
 &  & 8 & 5093.257(\;\:6) & 18(5) & -23 & 209 & 1.247 &  96 & 4f$^4$ $^5$I  &   2 & 4f$^4$ $^3$K2  &   1 & 4f$^4$ $^3$K1 \\
4f$^4$ & $^5$F & 2 & 10773.962(13) & 4(3) & 26 & -1325 & 1.007 &  93 & 4f$^4$ $^5$F  &   3 & 4f$^4$ $^3$D1  &   1 & 4f$^4$ $^3$D2 \\
 &  & 3 & 11424.605(\;\:8) & 9(1) & 23 & -1300 & 1.244 &  95 & 4f$^4$ $^5$F  &   1 & 4f$^4$ $^3$D1  &   1 & 4f$^4$ $^3$F3 \\
 &  & 4 & 12181.327(\;\:7) & 13(2) & 32 & -1279 & 1.330 &  92 & 4f$^4$ $^5$F  &   3 & 4f$^4$ $^3$G2  &   1 & 4f$^4$ $^3$F3 \\
 &  & 5 & 13210.278(\;\:8) & 10(4) & 41 & -1234 & 1.387 &  93 & 4f$^4$ $^5$F  &   3 & 4f$^4$ $^3$G2  &   1 & 4f$^4$ $^3$G1 \\
4f$^4$ & $^3$K2 & 6 & 14064.277(\;\:6) & 8(3) & -26 & -1002 & 0.860 &  78 & 4f$^4$ $^3$K2  &  20 & 4f$^4$ $^3$K1  &  &   \\
 &  & 7 & 15153.807(\;\:7) & 11(3) & 2 & -1027 & 1.005 &  69 & 4f$^4$ $^3$K2  &  18 & 4f$^4$ $^3$K1  &  11 & 4f$^4$ $^3$L \\
 &  & 8 & 16459.136(\;\:8) & 8(5) & 66 & -1007 & 1.105 &  63 & 4f$^4$ $^3$K2  &  14 & 4f$^4$ $^3$K1  &  11 & 4f$^4$ $^3$L \\
4f$^4$ & $^5$G & 3 & 15349.601(\;\:6) & 5(3) & -1 & -2018 & 0.885 &  73 & 4f$^4$ $^5$G  &  14 & 4f$^4$ $^3$G2  &   7 & 4f$^4$ $^3$G3 \\
 &  & 4 & 16831.344(10) & 5(3) & -75 & -1944 & 1.051 &  69 & 4f$^4$ $^5$G  &  13 & 4f$^4$ $^3$H4  &  10 & 4f$^4$ $^3$H3 \\
 &  & 5 & 18313.004(\;\:6) & 7(3) & -26 & -1825 & 1.167 &  57 & 4f$^4$ $^5$G  &  18 & 4f$^4$ $^3$H4  &  17 & 4f$^4$ $^3$H3 \\
4f$^4$ & $^3$H4 & 5 & 16622.337(\;\:7) & 7(1) & -11 & -1316 & 1.157 &  36 & 4f$^4$ $^5$G  &  21 & 4f$^4$ $^3$H4  &  13 & 4f$^4$ $^3$H3 \\
4f$^3$($^4$I)5d & $^5$L$^{\circ}$ & 6 & 15158.154(\;\:8) & 2(0) & -52 & -100 & 0.724 &  93 & ($^4$I)5d $^5$L  &   3 & ($^2$H2)5d $^3$K  &   3 & ($^4$I)5d $^3$K \\
 &  & 7 & 16952.835(\;\:8) & 2(0) & -29 & -32 & 0.914 &  97 & ($^4$I)5d $^5$L  &   2 & ($^2$H2)5d $^3$K  &   1 & ($^4$I)5d $^3$K \\
 &  & 8 & 18861.064(\;\:8) & 4(1) & -12 & 26 & 1.043 &  99 & ($^4$I)5d $^5$L  &   1 & ($^2$H2)5d $^3$K  &  &   \\
 &  & 9 & 20859.732(\;\:8) & 2(0) & 1 & 74 & 1.133 &  99 & ($^4$I)5d $^5$L  &  &    &  &   \\
 &  & 10 & 22932.276(11) & 1(0) & 11 & 61 & 1.200 &  99 & ($^4$I)5d $^5$L  &   1 & ($^2$K)5d $^3$M  &  &   \\
4f$^3$($^4$I)5d & $^5$K$^{\circ}$ & 5 & 15262.437(\;\:6) & 5(0) & 17 & 134 & 0.688 &  87 & ($^4$I)5d $^5$K  &   8 & ($^4$I)5d $^3$I  &   3 & ($^2$H2)5d $^3$I \\
 &  & 6 & 16938.068(\;\:8) & 5(1) & 36 & 217 & 0.912 &  94 & ($^4$I)5d $^5$K  &   4 & ($^4$I)5d $^3$I  &   2 & ($^2$H2)5d $^3$I \\
 &  & 7 & 18656.272(\;\:6) & 8(1) & 46 & 273 & 1.054 &  97 & ($^4$I)5d $^5$K  &   1 & ($^4$I)5d $^3$I  &   1 & ($^2$H2)5d $^3$I \\
 &  & 8 & 20410.897(\;\:6) & 7(0) & 49 & 306 & 1.150 &  97 & ($^4$I)5d $^5$K  &   1 & ($^4$I)5d $^3$L  &   1 & ($^2$K)5d $^3$L \\
 &  & 9 & 22197.115(\;\:6) & 5(1) & 44 & 315 & 1.218 &  96 & ($^4$I)5d $^5$K  &   3 & ($^4$I)5d $^3$L  &   2 & ($^2$K)5d $^3$L \\
4f$^3$($^4$I)5d & $^5$I$^{\circ}$ & 4 & 18883.669(\;\:5) & 5(0) & 76 & 170 & 0.634 &  83 & ($^4$I)5d $^5$I  &  13 & ($^4$I)5d $^3$H  &   1 & ($^4$G)5d $^3$H \\
 &  & 5 & 20348.742(\;\:5) & 7(0) & 50 & 140 & 0.920 &  74 & ($^4$I)5d $^5$I  &  12 & ($^4$I)5d $^3$H  &   5 & ($^4$I)5d $^3$I \\
 &  & 6 & 21980.522(\;\:6) & 5(1) & 24 & 136 & 1.081 &  68 & ($^4$I)5d $^5$I  &  11 & ($^4$I)5d $^3$I  &   8 & ($^4$I)5d $^3$H \\
 &  & 7 & 22702.777(\;\:6) & 12(0) & 36 & 204 & 1.150 &  58 & ($^4$I)5d $^5$I  &  21 & ($^4$I)5d $^3$I  &  12 & ($^4$I)5d $^3$K \\
 &  & 8 & 24592.259(\;\:6) & 7(0) & 16 & 291 & 1.231 &  85 & ($^4$I)5d $^5$I  &  12 & ($^4$I)5d $^3$K  &   2 & ($^2$K)5d $^3$K \\
4f$^3$($^4$I)5d & $^5$H$^{\circ}$ & 3 & 19593.094(\;\:7) & 3(1) & -15 & -160 & 0.591 &  62 & ($^4$I)5d $^5$H  &  22 & ($^4$I)5d $^3$G  &   5 & ($^4$F)5d $^5$H \\
 &  & 4 & 20388.995(\;\:5) & 6(0) & -67 & -76 & 0.836 &  43 & ($^4$I)5d $^5$H  &  32 & ($^4$I)5d $^3$H  &   9 & ($^4$I)5d $^5$I \\
 &  & 5 & 22181.299(12) & 4(1) & -43 & 62 & 1.035 &  56 & ($^4$I)5d $^5$H  &  16 & ($^4$I)5d $^3$I  &  11 & ($^4$I)5d $^3$H \\
 &  & 6 & 24058.518(\;\:6) & 5(0) & -48 & 66 & 1.128 &  50 & ($^4$I)5d $^5$H  &  26 & ($^4$I)5d $^3$I  &   5 & ($^4$I)5d $^3$K \\
 &  & 7 & 26097.906(\;\:6) & 7(2) & -56 & 76 & 1.193 &  46 & ($^4$I)5d $^5$H  &  28 & ($^4$I)5d $^3$I  &  13 & ($^4$I)5d $^3$K \\
4f$^3$($^4$I)5d & $^3$I$^{\circ}$ & 5 & 19616.622(\;\:6) & 5(0) & -17 & -5 & 0.865 &  54 & ($^4$I)5d $^3$I  &  17 & ($^4$I)5d $^5$I  &   9 & ($^4$I)5d $^5$K \\
 &  & 6 & 20799.349(\;\:6) & 8(3) & 0 & 22 & 1.026 &  40 & ($^4$I)5d $^3$I  &  26 & ($^4$I)5d $^5$I  &  11 & ($^4$I)5d $^3$K \\
 &  & 7 & 24003.225(\;\:6) & 8(1) & 10 & 61 & 1.171 &  37 & ($^4$I)5d $^3$I  &  33 & ($^4$I)5d $^5$I  &  16 & ($^4$I)5d $^5$H \\
4f$^3$($^4$I)5d & $^5$G$^{\circ}$ & 2 & 21477.328(\;\:9) & 1(0) & 7 & -9 & 0.344 &  93 & ($^4$I)5d $^5$G  &   3 & ($^2$H2)5d $^3$F  &   3 & ($^4$G)5d $^5$G \\
 &  & 3 & 21479.404(\;\:5) & 4(0) & 1 & -112 & 0.800 &  54 & ($^4$I)5d $^5$G  &  20 & ($^4$I)5d $^5$H  &  19 & ($^4$I)5d $^3$G \\
 &  & 4 & 23010.450(11) & 5(2) & -2 & -55 & 1.030 &  55 & ($^4$I)5d $^5$G  &  17 & ($^4$I)5d $^3$H  &  12 & ($^4$I)5d $^5$H \\
 &  & 5 & 24878.412(11) & 5(1) & -14 & -61 & 1.148 &  47 & ($^4$I)5d $^5$G  &  29 & ($^4$I)5d $^3$H  &   7 & ($^4$I)5d $^5$H \\
 &  & 6 & 26670.461(\;\:9) & 6(1) & -29 & -150 & 1.243 &  48 & ($^4$I)5d $^5$G  &  26 & ($^4$I)5d $^3$H  &   6 & ($^4$G)5d $^3$H \\
4f$^3$($^4$I)5d & $^3$H$^{\circ}$ & 4 & 21491.882(\;\:6) & 6(1) & 11 & -201 & 0.952 &  31 & ($^4$I)5d $^5$H  &  20 & ($^4$I)5d $^3$G  &  19 & ($^4$I)5d $^5$G \\
 &  & 5 & 23347.721(\;\:6) & 5(0) & 14 & -126 & 1.149 &  37 & ($^4$I)5d $^5$G  &  20 & ($^4$I)5d $^5$H  &  18 & ($^4$I)5d $^3$H \\
 &  & 6 & 25034.832(\;\:8) & 6(3) & -30 & -127 & 1.244 &  43 & ($^4$I)5d $^5$G  &  23 & ($^4$I)5d $^3$H  &  16 & ($^4$I)5d $^5$H \\
4f$^3$($^4$I)5d & $^3$K$^{\circ}$ & 6 & 23120.073(\;\:6) & 7(1) & -37 & -52 & 0.931 &  57 & ($^4$I)5d $^3$K  &  12 & ($^4$I)5d $^5$H  &   8 & ($^2$H2)5d $^3$K \\
 &  & 7 & 25319.579(\;\:7) & 5(2) & -21 & 21 & 1.096 &  51 & ($^4$I)5d $^3$K  &  25 & ($^4$I)5d $^5$H  &   8 & ($^2$H2)5d $^3$K \\
 &  & 8 & 27441.911(\;\:6) & 5(0) & 14 & -23 & 1.138 &  69 & ($^4$I)5d $^3$K  &  12 & ($^4$I)5d $^5$I  &  11 & ($^2$H2)5d $^3$K \\
4f$^3$($^4$I)5d & $^3$G$^{\circ}$ & 3 & 23891.226(\;\:6) & 3(0) & 11 & -300 & 0.796 &  48 & ($^4$I)5d $^3$G  &  36 & ($^4$I)5d $^5$G  &   8 & ($^4$I)5d $^5$H \\
 &  & 4 & 26140.225(10) & 4(2) & 10 & -382 & 1.060 &  66 & ($^4$I)5d $^3$G  &  20 & ($^4$I)5d $^5$G  &   5 & ($^4$I)5d $^5$H \\
 &  & 5 & 28440.047(\;\:7) & 3(0) & -4 & -489 & 1.194 &  77 & ($^4$I)5d $^3$G  &   8 & ($^4$I)5d $^5$G  &   3 & ($^4$F)5d $^3$G \\
4f$^3$($^4$I)5d & $^3$L$^{\circ}$ & 7 & 24497.165(\;\:7) & 4(1) & -29 & -454 & 0.891 &  82 & ($^4$I)5d $^3$L  &   7 & ($^2$K)5d $^3$L  &   3 & ($^4$I)5d $^3$K \\
 &  & 8 & 26676.685(\;\:9) & 5,(1) & 1 & -443 & 1.019 &  85 & ($^4$I)5d $^3$L  &   7 & ($^2$K)5d $^3$L  &   2 & ($^4$I)5d $^5$K \\
 &  & 9 & 28877.321(\;\:8) & 3(0) & 21 & -454 & 1.113 &  84 & ($^4$I)5d $^3$L  &   8 & ($^2$K)5d $^3$L  &   4 & ($^4$I)5d $^5$K \\
4f$^3$($^4$F)5d & $^5$H$^{\circ}$ & 3 & 27675.024(\;\:6) & 1(0) & -55 & -1819 & 0.589 &  77 & ($^4$F)5d $^5$H  &   6 & ($^4$I)5d $^5$H  &   4 & ($^2$D1)5d $^3$G \\
 &  & 4 & 28745.307(\;\:6) & 2(1) & -54 & -1761 & 0.907 &  86 & ($^4$F)5d $^5$H  &   6 & ($^4$I)5d $^5$H  &   2 & ($^2$D1)5d $^3$G \\
 &  & 5 & 30175.737(\;\:6) & 7(0) & -52 & -1676 & 1.040 &  52 & ($^4$F)5d $^5$H  &  10 & ($^2$H2)5d $^3$I  &   5 & ($^4$I)6s $^5$I \\
 &  & 6 & 31146.457(\;\:6) & 4(1) & -50 & -1494 & 1.111 &  56 & ($^4$F)5d $^5$H  &  18 & ($^2$H2)5d $^3$K  &   5 & ($^4$I)5d $^5$H \\
 &  & 7 & 32832.447(\;\:6) & 5(2) & -17 & -1401 & 1.142 &  41 & ($^4$F)5d $^5$H  &  33 & ($^2$H2)5d $^3$K  &   7 & ($^2$K)5d $^3$K \\
4f$^3$($^4$F)5d & $^5$D$^{\circ}$ & 3 & 28118.535(\;\:5) & 2(0) & 60 & -1800 & 1.297 &  26 & ($^4$F)5d $^5$D  &  24 & ($^4$S)5d $^5$D  &   9 & ($^4$F)5d $^5$H \\
4f$^3$($^4$F)5d & $^5$F$^{\circ}$ & 2 & 31888.511(\;\:9) & 3(1) & -41 & -2129 & 1.057 &  75 & ($^4$F)5d $^5$F  &   6 & ($^4$F)5d $^3$P  &   5 & ($^4$S)5d $^3$D \\
 &  & 3 & 32489.101(\;\:8) & 5(4) & -12 & -2131 & 1.227 &  69 & ($^4$F)5d $^5$F  &   7 & ($^4$S)5d $^3$D  &   3 & ($^4$F)5d $^5$D \\
 &  & 4 & 33603.488(\;\:7) & 6(0) & 60 & -2264 & 1.121 &  48 & ($^4$F)5d $^5$F  &  18 & ($^2$H2)5d $^3$H  &   6 & ($^2$G1)5d $^3$H \\
 &  & 5 & 34511.186(\;\:8) & 4(1) & -35 & -2116 & 1.237 &  54 & ($^4$F)5d $^5$F  &  13 & ($^2$H2)5d $^1$H  &   7 & ($^4$G)5d $^5$I \\
4f$^3$($^4$F)5d & $^3$H$^{\circ}$ & 4 & 32077.113(\;\:7) & 3(0) & 11 & -1678 & 0.950 &  12 & ($^4$F)5d $^3$H  &  12 & ($^2$H2)5d $^1$G  &  10 & ($^2$G1)5d $^3$H \\
 &  & 6 & 35154.437(\;\:7) & 4(0) & 56 & -1536 & 1.116 &  21 & ($^4$G)5d $^5$I  &  20 & ($^4$F)5d $^3$H  &   9 & ($^2$H2)5d $^3$H \\
4f$^3$($^4$F)5d & $^5$G$^{\circ}$ & 6 & 33210.660(13) & 5(3) & -36 & -1579 & 1.129 &  34 & ($^2$H2)5d $^3$I  &  34 & ($^4$F)5d $^5$G  &   7 & ($^2$H1)5d $^3$I \\
4f$^3$($^2$H2)5d & $^3$I$^{\circ}$ & 5 & 29397.423(\;\:6) & 6(1) & -113 & -1501 & 0.995 &  32 & ($^4$F)5d $^5$H  &  16 & ($^2$H2)5d $^3$I  &   7 & ($^2$G1)5d $^3$I \\
 &  & 7 & 35188.278(13) & 4(2) & -9 & -969 & 1.136 &  54 & ($^2$H2)5d $^3$I  &  13 & ($^2$H1)5d $^3$I  &   4 & ($^2$G1)5d $^3$I \\
4f$^3$($^2$H2)5d & $^3$K$^{\circ}$ & 6 & 29643.186(\;\:6) & 5(1) & 8 & -1212 & 0.994 &  22 & ($^2$H2)5d $^3$K  &  15 & ($^4$I)5d $^3$K  &   8 & ($^2$H2)5d $^1$I \\
 &  & 7 & 31781.802(\;\:8) & 5(1) & 9 & -1352 & 1.154 &  37 & ($^4$F)5d $^5$H  &  23 & ($^2$H2)5d $^3$K  &   9 & ($^2$G1)5d $^3$I \\
 &  & 8 & 34534.549(\;\:8) & 3(0) & -0 & -1188 & 1.129 &  58 & ($^2$H2)5d $^3$K  &  12 & ($^2$K)5d $^3$K  &   9 & ($^4$I)5d $^3$K \\
4f$^3$($^2$H2)5d & $^1$H$^{\circ}$ & 5 & 32605.894(\;\:8) & 10(6) & 37 & -1372 & 1.071 &  14 & ($^2$H2)5d $^1$H  &  11 & ($^4$F)5d $^5$F  &   9 & ($^2$H2)5d $^3$H \\
4f$^3$($^2$H2)5d & $^1$I$^{\circ}$ & 6 & 32823.701(\;\:8) & 3(0) & 32 & -1331 & 1.079 &  26 & ($^2$H2)5d $^1$I  &  19 & ($^4$F)5d $^5$G  &  13 & ($^2$H2)5d $^3$I \\
4f$^3$($^2$H2)5d & $^3$G$^{\circ}$ & 4 & 33135.917(\;\:7) & 6(1) & -38 & -2324 & 1.043 &  27 & ($^4$F)5d $^5$F  &  14 & ($^2$H2)5d $^3$G  &  13 & ($^4$G)5d $^5$I \\
4f$^3$($^2$H2)5d & $^3$H$^{\circ}$ & 5 & 34994.015(\;\:7) & 5(1) & -22 & -1995 & 1.100 &  23 & ($^2$H2)5d $^3$H  &  11 & ($^4$F)5d $^3$H  &  10 & ($^2$H2)5d $^3$G \\
4f$^3$($^2$H2)5d & $^1$G$^{\circ}$ & 4 & 36517.350(\;\:6) & 6(3) & -36 & -2641 & 1.054 &  32 & ($^4$G)5d $^5$G  &  26 & ($^4$G)5d $^5$H  &   9 & ($^2$H2)5d $^1$G \\
4f$^3$($^2$G1)5d & $^3$H$^{\circ}$ & 6 & 31559.217(\;\:6) & 6(3) & 48 & -1402 & 1.120 &  20 & ($^4$F)5d $^5$G  &  18 & ($^4$F)5d $^5$H  &  16 & ($^2$H2)5d $^3$K \\
4f$^3$($^2$G1)5d & $^3$G$^{\circ}$ & 3 & 34856.862(\;\:6) & 3(2) & -20 & -2332 & 0.780 &  29 & ($^4$G)5d $^5$H  &  16 & ($^2$G1)5d $^3$G  &  15 & ($^4$G)5d $^5$G \\
4f$^3$($^4$S)5d & $^5$D$^{\circ}$ & 1 & 32374.355(14) & 1(0) & -15 & -1850 & 1.036 &  37 & ($^4$S)5d $^5$D  &  12 & ($^4$F)5d $^5$D  &  11 & ($^4$S)5d $^3$D \\
 &  & 2 & 32905.369(\;\:8) & 2(1) & 24 & -1671 & 1.322 &  37 & ($^4$S)5d $^5$D  &  29 & ($^4$F)5d $^5$D  &  11 & ($^2$H2)5d $^3$F \\
 &  & 3 & 33400.822(\;\:7) & 2(0) & 23 & -1705 & 1.374 &  33 & ($^4$S)5d $^5$D  &  29 & ($^4$F)5d $^5$D  &   7 & ($^4$G)5d $^5$D \\
 &  & 4 & 34419.877(\;\:8) & 2(0) & 65 & -1826 & 1.352 &  29 & ($^2$H2)5d $^3$F  &  27 & ($^4$S)5d $^5$D  &  15 & ($^4$F)5d $^5$D \\
4f$^3$($^4$G)5d & $^5$I$^{\circ}$ & 4 & 32829.311(\;\:7) & 2(0) & 1 & -2053 & 0.786 &  61 & ($^4$G)5d $^5$I  &   6 & ($^4$F)5d $^3$F  &   5 & ($^4$G)5d $^3$F \\
 &  & 5 & 34017.306(\;\:6) & 4(1) & 36 & -2323 & 0.944 &  76 & ($^4$G)5d $^5$I  &   8 & ($^4$F)5d $^5$F  &   5 & ($^2$H2)5d $^3$I \\
 &  & 6 & 35549.565(\;\:6) & 3(0) & 16 & -2719 & 1.100 &  57 & ($^4$G)5d $^5$I  &  12 & ($^4$F)5d $^3$H  &   5 & ($^4$F)5d $^5$G \\
 &  & 7 & 36890.347(\;\:9) & 8(3) & 42 & -3010 & 1.146 &  44 & ($^4$G)5d $^5$I  &  43 & ($^4$I)6s $^3$I  &   4 & ($^2$K)5d $^3$K \\
 &  & 8 & 38517.773(\;\:7) & 3(1) & 14 & -2845 & 1.193 &  65 & ($^4$G)5d $^5$I  &  21 & ($^2$K)5d $^3$K  &   5 & ($^2$K)5d $^3$L \\
4f$^3$($^4$G)5d & $^5$H$^{\circ}$ & 3 & 35576.326(\;\:6) & 2(0) & 53 & -2129 & 0.692 &  48 & ($^4$G)5d $^5$H  &  25 & ($^4$G)5d $^5$G  &   9 & ($^4$F)5d $^3$G \\
 &  & 4 & 36235.179(\;\:6) & 3(1) & -29 & -1689 & 1.005 &  33 & ($^4$G)5d $^5$H  &   8 & ($^2$H2)5d $^3$F  &   7 & ($^2$H2)5d $^3$G \\
 &  & 5 & 37592.759(\;\:6) & 5(1) & -28 & -3092 & 1.109 &  67 & ($^4$G)5d $^5$H  &  14 & ($^4$G)5d $^5$G  &   4 & ($^4$F)5d $^3$H \\
 &  & 6 & 39058.274(\;\:6) & 4(0) & 68 & -2695 & 1.189 &  62 & ($^4$G)5d $^5$H  &  11 & ($^4$G)5d $^5$G  &   4 & ($^2$H2)5d $^3$I \\
 &  & 7 & 40505.082(\;\:6) & 3(1) & 134 & -2793 & 1.210 &  55 & ($^4$G)5d $^5$H  &  11 & ($^4$G)5d $^3$I  &  11 & ($^2$K)5d $^3$I \\
4f$^3$($^4$G)5d & $^5$G$^{\circ}$ & 4 & 36572.870(\;\:6) & 3(1) & -15 & -3016 & 1.069 &  35 & ($^4$G)5d $^5$G  &  23 & ($^4$G)5d $^5$H  &   6 & ($^2$G1)5d $^3$G \\
 &  & 5 & 37977.641(\;\:6) & 6(1) & -37 & -2649 & 1.208 &  50 & ($^4$G)5d $^5$G  &   8 & ($^4$G)5d $^5$H  &   4 & ($^2$H2)5d $^3$H \\
 &  & 6 & 39611.389(\;\:6) & 3(0) & -45 & -2897 & 1.223 &  45 & ($^4$G)5d $^5$G  &  18 & ($^4$G)5d $^5$H  &   5 & ($^2$K)5d $^1$I \\
4f$^3$($^4$G)5d & $^3$I$^{\circ}$ & 6 & 44609.232(\;\:7) & 2(0) & 21 & -3380 & 1.018 &  34 & ($^4$G)5d $^3$I  &  16 & ($^2$K)5d $^3$I  &   7 & ($^2$H1)5d $^3$I \\
 &  & 7 & 45592.778(\;\:9) & 1(0) & -28 & -3468 & 1.117 &  41 & ($^4$G)5d $^3$I  &  18 & ($^2$K)5d $^3$I  &   7 & ($^2$I)5d $^3$K \\
4f$^3$($^2$K)5d & $^3$I$^{\circ}$ & 7 & 41260.027(\;\:6) & 3(0) & 17 & -2715 & 1.173 &  49 & ($^2$K)5d $^3$I  &  26 & ($^4$G)5d $^5$H  &   8 & ($^4$G)5d $^3$I \\
4f$^3$($^4$D)5d & $^5$F$^{\circ}$ & 4 & 47404.355(\;\:9) & 3(0) & -35 & -5744 & 1.217 &  47 & ($^4$D)5d $^5$F  &  18 & ($^4$D)5d $^5$G  &   8 & ($^2$D1)5d $^1$G \\
4f$^3$($^4$I)6s & ($\frac{9}{2}$,$\frac{1}{2}$)$^{\circ}$ & 4 & 29873.222(\;\:8) & 5(1) & 15 & -1108 & 0.607 &  96 & ($^4$I)6s ($\frac{9}{2}$,$\frac{1}{2}$)  &   2 & ($^2$H)6s ($\frac{9}{2}$,$\frac{1}{2}$)  &   1 & ($^4$G)5d ($\frac{5}{2}$,$\frac{3}{2}$) \\
 &  & 5 & 30593.085(\;\:7) & 7(0) & 25 & -999 & 0.903 &  74 & ($^4$I)6s ($\frac{9}{2}$,$\frac{1}{2}$)  &  12 & ($^4$I)6s ($\frac{11}{2}$,$\frac{1}{2}$)  &   3 & ($^4$F)5d ($\frac{5}{2}$,$\frac{5}{2}$) \\
4f$^3$($^4$I)6s & ($\frac{11}{2}$,$\frac{1}{2}$)$^{\circ}$ & 6 & 32075.658(\;\:9) & 6(0) & -23 & -1056 & 1.063 &  90 & ($^4$I)6s ($\frac{11}{2}$,$\frac{1}{2}$)  &   6 & ($^4$I)6s ($\frac{13}{2}$,$\frac{1}{2}$)  &   1 & ($^2$H)5d ($\frac{11}{2}$,$\frac{3}{2}$) \\
 &  & 5 & 32309.729(10) & 7(4) & -64 & -930 & 0.873 &  81 & ($^4$I)6s ($\frac{11}{2}$,$\frac{1}{2}$)  &  11 & ($^4$I)6s ($\frac{9}{2}$,$\frac{1}{2}$)  &   1 & ($^2$H)5d ($\frac{9}{2}$,$\frac{5}{2}$) \\
4f$^3$($^4$I)6s & ($\frac{13}{2}$,$\frac{1}{2}$)$^{\circ}$ & 7 & 33906.143(10) & 6(1) & -2 & -1028 & 1.174 &  95 & ($^4$I)6s ($\frac{13}{2}$,$\frac{1}{2}$)  &   3 & ($^4$I)6s ($\frac{15}{2}$,$\frac{1}{2}$)  &  &   \\
 &  & 6 & 34687.230(\;\:9) & 4(0) & 28 & -993 & 1.035 &  85 & ($^4$I)6s ($\frac{13}{2}$,$\frac{1}{2}$)  &   7 & ($^4$I)6s ($\frac{11}{2}$,$\frac{1}{2}$)  &   3 & ($^4$G)5d ($\frac{9}{2}$,$\frac{3}{2}$) \\
4f$^3$($^4$I)6s & ($\frac{15}{2}$,$\frac{1}{2}$)$^{\circ}$ & 8 & 35839.828(12) & 5(2) & 7 & -1028 & 1.248 &  98 & ($^4$I)6s ($\frac{15}{2}$,$\frac{1}{2}$)  &   1 & ($^2$K)6s ($\frac{15}{2}$,$\frac{1}{2}$)  &  &   \\
 &  & 7 & 37089.599(\;\:7) & 8(3) & -3 & -832 & 1.148 &  50 & ($^4$I)6s ($\frac{15}{2}$,$\frac{1}{2}$)  &  27 & ($^4$G)5d ($\frac{9}{2}$,$\frac{5}{2}$)  &   8 & ($^4$G)5d ($\frac{11}{2}$,$\frac{3}{2}$) \\
4f$^3$($^4$I)6p & $^5$K & 5 & 60526.067(\;\:7) & 14(0) & 14 & 321 & 0.719 &  71 & ($^4$I)6p $^5$K  &  22 & ($^4$I)6p $^3$I  &   3 & ($^4$I)6p $^5$I \\
 &  & 6 & 62520.646(\;\:7) & 15(0) & 21 & 358 & 0.941 &  67 & ($^4$I)6p $^5$K  &  17 & ($^4$I)6p $^3$I  &   9 & ($^4$I)6p $^5$I \\
 &  & 7 & 64622.006(\;\:7) & 15(2) & 12 & 373 & 1.083 &  57 & ($^4$I)6p $^5$K  &  20 & ($^4$I)6p $^5$I  &  12 & ($^4$I)6p $^3$K \\
 &  & 8 & 68302.664(\;\:8) & 11(0) & 18 & 524 & 1.168 &  62 & ($^4$I)6p $^5$K  &  21 & ($^4$I)6p $^5$I  &  17 & ($^4$I)6p $^3$K \\
 &  & 9 & 70141.009(\;\:9) & 6(0) & 9 & 483 & 1.221 &  99 & ($^4$I)6p $^5$K  &   1 & ($^2$K)6p $^3$L  &  &   \\
4f$^3$($^4$I)6p & $^5$I & 4 & 60638.963(\;\:6) & 15(2) & -95 & 207 & 0.711 &  50 & ($^4$I)6p $^5$I  &  36 & ($^4$I)6p $^3$H  &   9 & ($^4$I)6p $^5$H \\
 &  & 5 & 62640.266(\;\:6) & 14(1) & 89 & 411 & 0.977 &  48 & ($^4$I)6p $^5$I  &  26 & ($^4$I)6p $^3$H  &  20 & ($^4$I)6p $^5$H \\
 &  & 6 & 64549.620(\;\:6) & 11(0) & 26 & 383 & 1.119 &  47 & ($^4$I)6p $^5$I  &  31 & ($^4$I)6p $^5$H  &  13 & ($^4$I)6p $^3$H \\
 &  & 7 & 66497.225(\;\:7) & 18(2) & 15 & 509 & 1.122 &  41 & ($^4$I)6p $^5$I  &  28 & ($^4$I)6p $^5$K  &  19 & ($^4$I)6p $^3$K \\
 &  & 8 & 66792.024(\;\:7) & 17(3) & -8 & 368 & 1.182 &  37 & ($^4$I)6p $^5$K  &  37 & ($^4$I)6p $^5$I  &  24 & ($^4$I)6p $^3$K \\
4f$^3$($^4$I)6p & $^5$H & 4 & 64243.914(\;\:7) & 17(1) & 10 & 494 & 0.752 &  42 & ($^4$I)6p $^5$I  &  32 & ($^4$I)6p $^5$H  &  23 & ($^4$I)6p $^3$H \\
 &  & 3 & 64347.699(\;\:8) & 9(0) & -26 & 443 & 0.507 &  96 & ($^4$I)6p $^5$H  &   3 & ($^2$H2)6p $^3$G  &   1 & ($^4$G)6p $^5$H \\
 &  & 6 & 65023.633(\;\:7) & 17(1) & -7 & 542 & 0.950 &  40 & ($^4$I)6p $^3$K  &  24 & ($^4$I)6p $^5$K  &  16 & ($^4$I)6p $^3$I \\
 &  & 5 & 66272.802(\;\:8) & 12(0) & 11 & 562 & 0.954 &  37 & ($^4$I)6p $^5$H  &  34 & ($^4$I)6p $^3$I  &  18 & ($^4$I)6p $^5$I \\
 &  & 7 & 66717.513(\;\:7) & 18(2) & -9 & 379 & 1.191 &  49 & ($^4$I)6p $^5$H  &  27 & ($^4$I)6p $^3$I  &  12 & ($^4$I)6p $^5$K \\
4f$^3$($^4$I)6p & $^3$I & 5 & 64350.812(\;\:7) & 18(0) & 16 & 525 & 0.850 &  36 & ($^4$I)6p $^3$I  &  25 & ($^4$I)6p $^5$I  &  22 & ($^4$I)6p $^5$K \\
 &  & 6 & 68347.584(\;\:8) & 14(2) & 2 & 557 & 1.091 &  49 & ($^4$I)6p $^3$I  &  36 & ($^4$I)6p $^5$H  &   8 & ($^4$I)6p $^5$I \\
 &  & 7 & 70460.602(\;\:8) & 14(3) & -14 & 548 & 1.186 &  59 & ($^4$I)6p $^3$I  &  32 & ($^4$I)6p $^5$H  &   3 & ($^4$I)6p $^5$I \\
4f$^3$($^4$I)6p & $^3$H & 4 & 66445.159(\;\:8) & 13(1) & 3 & 519 & 0.851 &  56 & ($^4$I)6p $^5$H  &  37 & ($^4$I)6p $^3$H  &   4 & ($^4$I)6p $^5$I \\
 &  & 5 & 68613.221(\;\:9) & 11(0) & 9 & 553 & 1.046 &  61 & ($^4$I)6p $^3$H  &  32 & ($^4$I)6p $^5$H  &   3 & ($^4$I)6p $^5$I \\
 &  & 6 & 70826.009(\;\:9) & 13(1) & -1 & 573 & 1.167 &  79 & ($^4$I)6p $^3$H  &  14 & ($^4$I)6p $^5$H  &   2 & ($^4$I)6p $^3$I \\
4f$^3$($^4$I)6p & $^3$K & 6 & 66451.812(\;\:7) & 14(1) & -4 & 575 & 0.973 &  47 & ($^4$I)6p $^3$K  &  28 & ($^4$I)6p $^5$I  &   9 & ($^4$I)6p $^5$H \\
 &  & 7 & 68507.387(\;\:8) & 13(0) & -11 & 592 & 1.094 &  55 & ($^4$I)6p $^3$K  &  34 & ($^4$I)6p $^5$I  &   7 & ($^4$I)6p $^5$H \\
 &  & 8 & 70599.570(\;\:8) & 16(0) & -23 & 591 & 1.176 &  57 & ($^4$I)6p $^3$K  &  41 & ($^4$I)6p $^5$I  &   1 & ($^2$K)6p $^1$L \\
4f$^3$($^4$F)6p & $^5$G & 2 & 72088.930(18) & 1(0) & -22 & -1939 & 0.427 &  77 & ($^4$F)6p $^5$G  &  14 & ($^4$F)6p $^3$F  &   4 & ($^2$D1)6p $^3$F \\
 &  & 3 & 73201.950(28) & 1(0) & -11 & -1919 & 0.947 &  73 & ($^4$F)6p $^5$G  &  12 & ($^4$F)6p $^3$F  &   6 & ($^4$F)6p $^3$G \\
 &  & 4 & 74258.973(23) & 2(0) & -8 & -1861 & 1.112 &  47 & ($^4$F)6p $^5$G  &  15 & ($^4$F)6p $^3$G  &   8 & 4f$^4$ $^1$G3 \\
 &  & 5 & 75579.962(31) & 1(0) & -41 & -1681 & 1.175 &  30 & ($^4$F)6p $^5$G  &  26 & ($^4$F)6p $^3$G  &  16 & ($^4$F)6p $^5$F \\
 &  & 6 & 78974.826(24) & 1(0) & -17 & -938 & 1.217 &  61 & ($^4$F)6p $^5$G  &  16 & ($^2$H2)6p $^1$I  &   9 & ($^2$H2)6p $^3$I \\
4f$^3$($^2$H2)6p & $^3$I & 5 & 73247.950(16) & 2(0) & 34 & -1068 & 0.961 &  39 & ($^2$H2)6p $^3$I  &  10 & ($^2$H2)6p $^1$H  &   6 & ($^2$H1)6p $^3$I \\
 &  & 6 & 76644.575(28) & 1(0) & -59 & -1037 & 1.053 &  54 & ($^2$H2)6p $^3$I  &   9 & ($^2$H1)6p $^3$I  &   8 & ($^2$H2)6p $^3$H \\
4f$^3$($^2$H2)6p & $^1$H & 5 & 77231.115(34) & 1(0) & -44 & -938 & 1.099 &  19 & ($^2$H2)6p $^1$H  &  18 & ($^4$F)6p $^5$G  &  11 & ($^2$H2)6p $^3$I \\
4f$^3$($^2$H2)6p & $^1$I & 6 & 77236.125(24) & 3(1) & 86 & -938 & 1.129 &  34 & ($^2$H2)6p $^1$I  &  19 & ($^4$F)6p $^5$G  &  16 & ($^2$H2)6p $^3$H \\
4f$^3$($^2$H2)6p & $^3$H & 6 & 80593.604(33) & 1(0) & 2 & -938 & 1.112 &  47 & ($^2$H2)6p $^3$H  &  21 & ($^2$H2)6p $^1$I  &   9 & ($^2$H1)6p $^3$H \\
\enddata
\tablecomments{The first 3 columns are the configuration, term and $J$ value assignments for the level, parent term of the 3 core electrons is in brackets, any number following a term symbol is used to distinguish recurrent terms of equivalent electrons and a $^{\circ}$ following a term symbol indicates a level of odd parity. The fourth column is the optimised energy and its associated uncertainty in brackets in units of $10^{-3}$~cm$^{-1}$, the fifth column lists the number of lines classified for each level in the Nd-Ar PDL FT spectra, the number in the bracket is the number of blended or weak lines omitted from energy level optimisation. The sixth and seventh columns are the differences between observed and calculated values of the present work using the parameterised Cowan code and of \cite{gaigalas2019extended} respectively. The remaining columns list the Land\'e $g$-factors and 3 leading eigenvector percentages calculated using the Cowan code in this work. The level energies are optimised using LOPT using all 580 classified transitions listed in Table \ref{tab:line_id}.
\\ (This table is available in machine-readable form)}
\end{deluxetable*}

Level energies from the parameterised Cowan code calculations of this work generally differ less than $50$~cm$^{-1}$ from the observed levels used to fit the radial integrals. This deviation is expected to be up to a few factors larger for the remaining experimentally unknown levels of the 4f$^4$, 4f$^3$5f, 4f$^3$6s and 4f$^3$6p configurations. In contrast, deviations of observed level energies from \cite{gaigalas2019extended} calculations were 1 or 2 orders larger in size but still on average within a few percent. Due to the increasing level density at higher energies, a few percent difference in higher-lying level energies may lead to changes in the calculated eigenvector compositions and transition probabilities which are key to the empirical spectrum analysis. For example, mixing between the 4f$^3$5d and 4f$^3$6s configurations was evident between the 4f\textsuperscript{3}(\textsuperscript{4}G)5d~\textsuperscript{5}I\textsubscript{7} and 4f\textsuperscript{3}($^4$I)6s $(\frac{15}{2},\frac{1}{2})_7$ levels. In \cite{gaigalas2019extended}, these two levels were predicted around 2000~cm$^{-1}$ apart, which was an order of magnitude larger than the observed energy difference. Transition probabilities of 4f\textsuperscript{3}(\textsuperscript{4}G)5d~\textsuperscript{5}I\textsubscript{7}~-~4f$^3$($^4$I)6p and 4f$^4$~-~4f\textsuperscript{3}($^4$I)6s $(\frac{15}{2},\frac{1}{2})_7$ from \cite{gaigalas2019extended} were up to almost 3 orders of magnitude lower compared to those calculated by the parameterised Cowan code, which were at the expected order of magnitude as found by comparison with the observed relative intensities.

\subsubsection{Transitions Classified in Laboratory Sources}\label{sec:lines}
All 433 Nd~III transitions originating from the 4f$^3$5d, 4f$^3$6s, and 4f$^3$6p configurations classified from the Nd-Ar PDL FT spectra are listed in Table \ref{tab:nd_ft_lines}. 
Around half of these transitions were measured with wavenumber uncertainties less than $0.01$~cm$^{-1}$, uncertainties above $0.02$~cm$^{-1}$ were from weak lines around the noise level and constituted no more than 25\% of this list.

All Nd~III lines below 3250~\AA{} classified in the Nd-Ar PDL FT spectra from Table \ref{tab:nd_ft_lines} were observed in the Nd VS grating spectra. Additionally, 191 of the 4f$^3$5d - 4f$^3$6p and 4f$^3$6s - 4f$^3$6p transitions with $g_uA<10^8$~s$^{-1}$ unobserved in the Nd-Ar PDL FT spectra were present in the Nd VS grating spectra, and these are listed in Table \ref{tab:nd_vs_lines}. For convenience, the $g_uA$ values are given in $10^6$~s$^{-1}$ to compare with the observed grating line intensities. These intensities were derived from the measured line blackening using model response curves of the photographic plates, which were placed onto a linear relative scale approximately corresponding to the $g_uA$ values. These intensities could be roughly compared with those from the FT spectra via a division of 100, but they should only be considered qualitative due to the uncertainty of the response curve of the grating spectrometers and the non-linearity of the photographic plate response. Nevertheless, the relative intensities generally followed the $g_uA$ values, which is remarkable for these weak lines originating from the overall highly mixed 4f$^3$6p levels.

\begin{longrotatetable}
\movetabledown=10mm
\begin{deluxetable*}{rrcrcccrcrcrcrrc}
\tabletypesize{\scriptsize}
\tablecaption{Classified transitions of Nd~III originating from the 4f$^3$5d, 4f$^3$6s, and 4f$^3$6p configurations in the Nd-Ar PDL FT spectra. \label{tab:nd_ft_lines}}
\colnumbers
\tablehead{
\colhead{SNR} & 
\colhead{Int.} & 
\colhead{$g_uA$} & 
\colhead{log($g_lf$)} & 
\colhead{FWHM} & 
\colhead{$\sigma$} & 
\colhead{$\sigma_{\text{Ritz}}$} & 
\colhead{$\sigma-\sigma_{\text{Ritz}}$} &
\colhead{$\lambda^{\text{air}}_{\text{Ritz}}$} & 
\multicolumn{2}{c}{Lower Level} &
\multicolumn{2}{c}{Upper Level} & 
\colhead{$E_l$} & 
\colhead{$E_u$} &
\colhead{Note}
\\
& 
\colhead{(arb.)} & 
\colhead{(s$^{-1}$)} & 
&
\colhead{(cm$^{-1}$)} &
\colhead{(cm$^{-1}$)} &
\colhead{(cm$^{-1}$)} & 
\colhead{(cm$^{-1}$)} & 
\colhead{(\AA)} & 
\colhead{Config.} & \colhead{Term$_J$} &
\colhead{Config.} & \colhead{Term$_J$} &
\colhead{(cm$^{-1}$)} & 
\colhead{(cm$^{-1}$)} & 
}
\startdata
10 & 20 & $1.3\times 10^7$ & -2.13 & 0.251 & 51599.527(28) & 51599.538(\;\:9) & -0.012 & - & 4f$^3$($^4$I)5d & $^5$L$_{8}$ & 4f$^3$($^4$I)6p & $^3$I$_{7}$ & 18861.064 & 70460.602 & B/W\\
5 & 11 & $6.1\times 10^8$ & -0.42 & 0.259 & 49192.654(29) & 49192.658(\;\:9) & -0.004 & 2032.1699 & 4f$^3$($^4$I)5d & $^5$L$_{6}$ & 4f$^3$($^4$I)6p & $^3$I$_{5}$ & 15158.154 & 64350.812 &  \\
8 & 6 & $5.0\times 10^8$ & -0.51 & 0.177 & 48981.473(32) & 48981.477(\;\:7) & -0.004 & 2040.9327 & 4f$^3$($^4$I)5d & $^5$K$_{5}$ & 4f$^3$($^4$I)6p & $^5$H$_{4}$ & 15262.437 & 64243.914 &  \\
5 & 5 & $7.4\times 10^8$ & -0.33 & 0.158 & 48402.467(24) & 48402.455(\;\:8) & 0.012 & 2065.3510 & 4f$^3$($^4$I)5d & $^5$K$_{9}$ & 4f$^3$($^4$I)6p & $^3$K$_{8}$ & 22197.115 & 70599.570 &  \\
6 & 7 & $6.5\times 10^8$ & -0.37 & 0.179 & 48096.502(21) & 48096.490(\;\:8) & 0.012 & 2078.4914 & 4f$^3$($^4$I)5d & $^5$K$_{8}$ & 4f$^3$($^4$I)6p & $^3$K$_{7}$ & 20410.897 & 68507.387 &  \\
13 & 15 & $9.2\times 10^8$ & -0.22 & 0.159 & 48070.793(10) & 48070.798(\;\:8) & -0.005 & 2079.6024 & 4f$^3$($^4$I)5d & $^5$L$_{7}$ & 4f$^3$($^4$I)6p & $^5$H$_{6}$ & 16952.835 & 65023.633 &  \\
13 & 18 & $2.5\times 10^9$ & 0.22 & 0.193 & 47943.883(10) & 47943.894(\;\:9) & -0.011 & 2085.1077 & 4f$^3$($^4$I)5d & $^5$K$_{9}$ & 4f$^3$($^4$I)6p & $^5$K$_{9}$ & 22197.115 & 70141.009 &  \\
8 & 11 & $1.1\times 10^9$ & -0.15 & 0.221 & 47891.776(17) & 47891.766(\;\:8) & 0.010 & 2087.3775 & 4f$^3$($^4$I)5d & $^5$K$_{8}$ & 4f$^3$($^4$I)6p & $^5$K$_{8}$ & 20410.897 & 68302.664 &  \\
7 & 7 & $4.8\times 10^8$ & -0.50 & 0.166 & 47856.459(17) & 47856.449(\;\:8) & 0.010 & 2088.9181 & 4f$^3$($^4$I)5d & $^5$L$_{8}$ & 4f$^3$($^4$I)6p & $^5$H$_{7}$ & 18861.064 & 66717.513 &  \\
3 & 5 & $2.8\times 10^8$ & -0.74 & 0.275 & 47840.961(53) & 47840.953(\;\:7) & 0.009 & 2089.5948 & 4f$^3$($^4$I)5d & $^5$K$_{7}$ & 4f$^3$($^4$I)6p & $^5$I$_{7}$ & 18656.272 & 66497.225 &  \\
7 & 9 & $5.7\times 10^8$ & -0.43 & 0.233 & 47795.548(20) & 47795.539(\;\:7) & 0.009 & 2091.5805 & 4f$^3$($^4$I)5d & $^5$K$_{7}$ & 4f$^3$($^4$I)6p & $^3$K$_{6}$ & 18656.272 & 66451.812 &  \\
15 & 23 & $1.4\times 10^9$ & -0.02 & 0.214 & 47636.162(10) & 47636.161(\;\:8) & 0.001 & 2098.5793 & 4f$^3$($^4$I)5d & $^5$L$_{8}$ & 4f$^3$($^4$I)6p & $^5$I$_{7}$ & 18861.064 & 66497.225 &  \\
26 & 32 & $3.3\times 10^9$ & 0.35 & 0.224 & 47442.932(\;\:7) & 47442.932(\;\:8) & -0.000 & 2107.1276 & 4f$^3$($^4$I)5d & $^5$L$_{9}$ & 4f$^3$($^4$I)6p & $^5$K$_{8}$ & 20859.732 & 68302.664 &  \\
$\vdots$ & $\vdots$ & $\vdots$ & $\vdots$ & $\vdots$ & $\vdots$ & $\vdots$ & $\vdots\quad$ & $\vdots$ & $\vdots\quad\quad$ & $\vdots$ & $\vdots\quad\quad$ & $\vdots$ & $\vdots\quad\quad$ & $\vdots\quad\quad$ & \\
15 & 152 & $7.1\times 10^6$ & -1.34 & 0.123 & 15322.669(10) & 15322.665(\;\:8) & 0.004 & 6524.4772 & 4f$^4$ & $^3$K$_{8}$2 & 4f$^3$($^2$H2)5d & $^3$K$_{7}$ & 16459.136 & 31781.802 &  \\
12 & 134 & $5.9\times 10^5$ & -2.42 & 0.128 & 15317.632(13) & 15317.640(\;\:6) & -0.008 & 6526.6176 & 4f$^4$ & $^5$I$_{8}$ & 4f$^3$($^4$I)5d & $^5$K$_{8}$ & 5093.257 & 20410.897 &  \\
312 & 3428 & $5.8\times 10^6$ & -1.43 & 0.178 & 15262.437(\;\:2) & 15262.437(\;\:6) & -0.000 & 6550.2239 & 4f$^4$ & $^5$I$_{4}$ & 4f$^3$($^4$I)5d & $^5$K$_{5}$ & 0.000 & 15262.437 &  \\
12 & 137 & $6.3\times 10^5$ & -2.37 & 0.141 & 14941.721(13) & 14941.725(\;\:6) & -0.004 & 6690.8204 & 4f$^4$ & $^5$I$_{7}$ & 4f$^3$($^4$I)5d & $^5$K$_{7}$ & 3714.548 & 18656.272 &  \\
5 & 37 & $2.3\times 10^6$ & -1.81 & 0.089 & 14822.931(26) & 14822.913(\;\:7) & 0.019 & 6744.4507 & 4f$^4$ & $^5$G$_{5}$ & 4f$^3$($^2$H2)5d & $^3$G$_{4}$ & 18313.004 & 33135.917 & B/W\\
13 & 170 & $1.9\times 10^6$ & -1.88 & 0.145 & 14715.629(12) & 14715.620(10) & 0.009 & 6793.6254 & 4f$^4$ & $^5$F$_{3}$ & 4f$^3$($^4$I)5d & $^3$G$_{4}$ & 11424.605 & 26140.225 &  \\
12 & 175 & $4.5\times 10^5$ & -2.50 & 0.136 & 14550.517(13) & 14550.524(\;\:8) & -0.007 & 6870.7090 & 4f$^4$ & $^5$I$_{6}$ & 4f$^3$($^4$I)5d & $^5$K$_{6}$ & 2387.544 & 16938.068 &  \\
6 & 99 & $3.2\times 10^6$ & -1.64 & 0.151 & 14489.386(13) & 14489.378(\;\:7) & 0.007 & 6899.7039 & 4f$^4$ & $^3$K$_{7}$2 & 4f$^3$($^2$H2)5d & $^3$K$_{6}$ & 15153.807 & 29643.186 &  \\
4 & 72 & $2.3\times 10^5$ & -2.77 & 0.134 & 14124.631(17) & 14124.644(\;\:6) & -0.013 & 7077.8731 & 4f$^4$ & $^5$I$_{5}$ & 4f$^3$($^4$I)5d & $^5$K$_{5}$ & 1137.794 & 15262.437 &  \\
3 & 79 & $1.3 \times 10^3$ & -3.73 & 0.122 & 13562.995(22) & 13563.015(\;\:6) & -0.020 & 7370.9613 & 4f$^4$ & $^5$I$_{8}$ & 4f$^3$($^4$I)5d & $^5$K$_{7}$ & 5093.257 & 18656.272 & B/W\\
11 & 421 & $6.3\times 10^6$ & -1.29 & 0.130 & 13460.188(15) & 13460.183(10) & 0.005 & 7427.2739 & 4f$^4$ & $^5$F$_{5}$ & 4f$^3$($^4$I)5d & $^5$G$_{6}$ & 13210.278 & 26670.461 &  \\
6 & 197 & $1.8\times 10^6$ & -1.81 & 0.110 & 13117.264(12) & 13117.264(13) & 0.000 & 7621.4429 & 4f$^4$ & $^5$F$_{2}$ & 4f$^3$($^4$I)5d & $^3$G$_{3}$ & 10773.962 & 23891.226 &  \\
4 & 340 & $3.8\times 10^6$ & -1.45 & 0.187 & 12697.081(22) & 12697.085(11) & -0.004 & 7873.6571 & 4f$^4$ & $^5$F$_{4}$ & 4f$^3$($^4$I)5d & $^5$G$_{5}$ & 12181.327 & 24878.412 &  \\
\enddata
\tablecomments{The columns are: (1) signal-to-noise ratio, (2) approximate relative intensity, proportional to the photon rate, large uncertainties may apply when comparing intensities between the six spectra listed in Table \ref{tab:spec_params} due to different lamp conditions, (3)-(4) weighted transition probability and log of the weighted (absorption) oscillator strength calculated using the Cowan code, where $g_u$ and $g_l$ refer to statistical weights of the upper and lower energy levels respectively, (5) full width at half maximum of the fitted line, (6)-(7) observed wavenumber and Ritz wavenumber optimised from energy level fitting, their uncertainties are given in brackets in units of ($10^{-3}$~cm$^{-1}$), (8) wavenumber difference between observed and Ritz values, (9) Ritz air wavelength ($>$200~nm) converted using the refractive index of air from \cite{peck1972dispersion}, (10)-(13) energy levels associated with the transition, their energies are in columns (14)-(15) respectively, and (16) contains comments of the observed transition, where B/W indicates blended or weak line with unreliable wavenumber and intensity, which was omitted from energy level fitting.
\\(The full version of this table is available in machine-readable form.)}
\end{deluxetable*}
\end{longrotatetable}

\begin{deluxetable*}{rrccrccllcc}
\tabletypesize{\scriptsize}
\tablecaption{Transitions of Nd~III originating from the 4f$^3$6s and 4f$^3$6p configurations observed only in the Nd VS grating spectra. \label{tab:nd_vs_lines}}
\tabletypesize{\scriptsize}
\colnumbers
\tablehead{
\colhead{Int.} & 
\colhead{$g_uA$} & 
\colhead{log($g_lf$)} & 
\colhead{$\lambda$} &
\colhead{$\lambda - \lambda_{\text{Ritz}}$} & 
\colhead{$\lambda^{\text{air}}$} &
\colhead{$\sigma$} & 
\colhead{Lower Level} &
\colhead{Upper Level} & 
\colhead{$E_l$} & 
\colhead{$E_u$} 
\\
& 
\colhead{($10^{6}$~s$^{-1}$)} & 
&
\colhead{(\AA)} & 
\colhead{(\AA)} &
\colhead{(\AA)} &
\colhead{(cm$^{-1}$)} &
&
&
\colhead{(cm$^{-1}$)} & 
\colhead{(cm$^{-1}$)} 
}
\startdata
 315 &  161 & -1.06 & 1904.916 &  0.003 & 1904.916 & 52495.75 & 4f$^3$($^4$I)5d $^5$H$_{3}$ & 4f$^3$($^4$F)6p $^5$G$_{2}$ & 19593.094 & 72088.930\\
 388 &  205 & -0.95 & 1906.218 & -0.001 & 1906.218 & 52459.91 & 4f$^3$($^4$I)5d $^3$K$_{6}$ & 4f$^3$($^4$F)6p $^5$G$_{5}$ & 23120.073 & 75579.962\\
  65 &   65 & -1.44 & 1933.861 &  0.001 & 1933.861 & 51710.03 & 4f$^3$($^4$I)5d $^3$H$_{4}$ & 4f$^3$($^4$F)6p $^5$G$_{3}$ & 21491.882 & 73201.950\\
 103 &   55 & -1.50 & 1950.555 & -0.002 & 1950.555 & 51267.47 & 4f$^3$($^4$I)5d $^5$I$_{6}$ & 4f$^3$($^2$H2)6p $^3$I$_{5}$ & 21980.522 & 73247.948\\
 114 &   97 & -1.25 & 1977.631 &  0.004 & 1977.631 & 50565.56 & 4f$^3$($^4$I)5d $^5$G$_{6}$ & 4f$^3$($^2$H2)6p $^1$I$_{6}$ & 26670.462 & 77236.122\\
 $\vdots\:$ & $\vdots\:$ & $\vdots$ & $\vdots$ & $\vdots\quad$ & $\vdots$ & $\vdots$ & $\quad\quad\quad\vdots$ & $\quad\quad\quad\vdots$ & $\vdots$ & $\vdots$ \\
 760 &   44 & -1.17 & 3183.563 &  0.001 & 3182.643 & 31411.34 & 4f$^3$($^4$F)5d $^5$G$_{6}$ & 4f$^3$($^4$I)6p $^5$K$_{7}$ & 33210.659 & 64622.009\\
 631 &   32 & -1.31 & 3190.919 &  0.003 & 3189.996 & 31338.94 & 4f$^3$($^4$F)5d $^5$G$_{6}$ & 4f$^3$($^4$I)6p $^5$I$_{6}$ & 33210.659 & 64549.622\\
 501 &   34 & -1.28 & 3200.866 & -0.000 & 3199.942 & 31241.54 & 4f$^3$($^2$H2)5d $^3$I$_{5}$ & 4f$^3$($^4$I)6p $^5$I$_{4}$ & 29397.423 & 60638.962\\
 394 &   17 & -1.59 & 3212.476 &  0.001 & 3211.549 & 31128.63 & 4f$^3$($^2$H2)5d $^3$I$_{5}$ & 4f$^3$($^4$I)6p $^5$K$_{5}$ & 29397.423 & 60526.066\\
 692 &  186 & -0.51 & 3213.624 & -0.003 & 3212.696 & 31117.52 & 4f$^3$($^4$I)6s ($\frac{13}{2}$,$\frac{1}{2}$)$_{7}$ & 4f$^3$($^4$I)6p $^5$H$_{6}$ & 33906.143 & 65023.636\\
\enddata
\tablecomments{The columns are: (1) approximate relative intensity, proportional to the photon rate, (2)-(3) weighted transition probability and log of the weighted (absorption) oscillator strength calculated using the Cowan code, where $g_u$ and $g_l$ refer to statistical weights of the upper and lower energy levels respectively, (4) calibrated wavelength of the observed line, where uncertainties of unblended and symmetric lines of moderate intensity were estimated at 0.006~\AA{}, (5) difference between observed wavelength and Ritz wavelength optimised from energy level fitting, (6) air wavelength converted using column (3) and the refractive index of air from \cite{peck1972dispersion}, vacuum wavelengths are given for values under 2000~\AA, (7) observed and calibrated wavenumber (8)-(11) information of energy levels associated with the transition.
\\(The full version of this table is available in machine-readable form.)}
\end{deluxetable*}

\subsection{Methodology of Analysis}\label{sec:analysis_method}
The empirical spectrum analysis of complex atomic structures involves the classification of spectral lines belonging to the experimentally unknown energy levels of interest, primarily by matching theoretical and experimental relative intensity patterns at wavenumbers expected from theoretical level energies and energy separations. 

With accurate transition probability calculations, one can expect the observed relative line intensities to be similar to those calculated for the lines. For example, the eighth and final rows of the top half of Table \ref{tab:nd_ft_lines} are two lines at 47891.776~cm$^{-1}$ and 47442.932~cm$^{-1}$ originating from the upper level 4f$^3$($^4$I)6p~$^5$K$_8$, with observed relative intensities expected from their theoretical branching ratios of $g_uA$. Similarly, the final three rows of the bottom half of Table \ref{tab:nd_ft_lines} show good agreement between the ratios of their $g_uA$ values and the ratios of their observed relative intensities, as the three lines were all in the same spectrum and with upper levels at similar energies.

To begin the analysis, a few energy levels of the ground and lowest-lying configurations must be established for an atom. This usually involves classifying the corresponding transitions which tend to be of the highest intensities and/or are self-absorbed in the discharge conditions that best populate the energy levels of the atom. The classification of these transitions is also often approached by looking for repeated wavenumber separations of observed lines \citep[e.g., see the end of Chapter 1 of][]{cowan1981theory}. Then, the level system could be expanded level by level by matching the expected intensity pattern and wavenumbers of lines connecting an unknown level to the known levels. 

Due to the immense number of spectral lines observed from complex atoms, the number of ambiguous matches for a particular intensity pattern rises exponentially with increasing experimental and theoretical wavenumber uncertainties. This issue escalates further when looking for low-intensity lines, which tend to be much more common and have higher uncertainties in both wavenumber and intensity. The solution then is to find corroboration with other available experimental evidence; in the analysis of Nd~III, the correct lines for transitions connecting a potential new level should:
\begin{enumerate}
    \item match the energy differences between known levels and the level of interest, within a tolerance around the order of observed wavenumber uncertainties (0.05~cm$^{-1}$ was suitable in most cases as it was around the upper-bound of wavenumber uncertainties of the Nd-Ar PDL FT spectra line list),
    \item satisfy the E1 transition selection rules in the absence of external fields, i.e., change of parity and $\Delta J = 0\pm1$ but not $J=0\leftrightarrow0$,
    \item be within the expected wavenumber range, subject to the accuracy of theoretical level energy predictions,
    \item show agreement between observed and predicted relative intensity patterns,
    \item correspond to a tentative energy level which would enable the establishment of other experimentally unknown levels from classifying transitions connecting to this tentative level,
    \item have the expected isotope and/or hyperfine structure line profiles,
    \item have the expected relative intensities in different thermal conditions, e.g., stars, the cooler HCL or hotter VS discharges,
    \item have the expected Zeeman patterns in stellar spectra or other suitable light sources using theoretical Land\'{e} \textit{g}-factors,
    \item and not be lines from other atomic species.
\end{enumerate}
In most cases, not all of the above conditions could be satisfied; the first four were generally considered necessary and the rest served as evidence for more confident classifications, which were subject to data availability. An energy level could also be established with only one line if the said line satisfied many of these conditions without ambiguities. The first 3 conditions were written as a computer program \citep[similar to][]{azarov2018iden2} to filter all observed wavenumbers to generate candidate level energies and their corresponding sets of candidate lines. Newly established energy levels were also used to empirically improve the theoretical calculations as the analysis progressed, which reduced the wavenumber and relative intensity search ranges.

The described methodology greatly depends on the completeness, spectral range, and accuracy of the observed wavenumbers and relative intensities. From the discussion in Section \ref{sec:spectrum}, the ubiquity of blended and weak lines of Nd could prevent the detection of crucial transitions or force an increase of uncertainty tolerances, resulting in an unresolvable number of ambiguous matches. The present work eventually reached a conclusion due to this reason.

\subsection{Revision of Previously Published Energy Levels}
The analysis began with the assumption of the correct establishment of the ground term, 4f$^4$~$^5$I, by H. M. Crosswhite \citep{martin1978atomic}. This became more evident and was concluded indeed correct as the analysis progressed; the 5 levels of the 4f$^4$~$^5$I term enabled the identification of more than 200 inter-connected levels of the singly-excited configurations up to $nl=\text{5f}$, consistent with observed spectra, isotope structure line profiles, and parameterised Cowan code calculations.

The immediate search for the 24 levels of the 4f$^3$5d configuration from \cite{martin1978atomic} was aided by calculations from \cite{zhang2002measurement}, \cite{ryabchikova2006rare} and \cite{gaigalas2019extended}. Of the 35 4f$^3$5d levels proposed by \cite{ryabchikova2006rare} all were confirmed except for the 4f$^3$($^4$I)5d~$^5$H$_3$ level, which was revised using its transition to the ground level classified at 19593.094~cm$^{-1}$ and transitions from 4f$^3$($^4$I)6p~$^5$I$_4$ and 4f$^3$($^4$I)6p~$^5$H$_3$. A positive wavenumber shift of order 0.01~cm$^{-1}$ exists for most energy levels and transitions observed in this work compared to those of \cite{aldenius2001} and \cite{ryabchikova2006rare}. This was from the use of COG values to account for the observed Nd~III isotope line profiles, rather than the peak wavenumbers used by \cite{aldenius2001} which did not consider energy level shifts from nuclear perturbations.

\subsection{The Search for Previously Unknown Energy Levels}
\subsubsection{\texorpdfstring{The 4f$^3$($^4$I) sub-configuration}{The 4f3(4I) sub-configuration}}
Theoretical calculations from \cite{ryabchikova2006rare} and \cite{gaigalas2019extended} aided the complete identification of all 40 levels of the 4f$^3$($^4$I)5d, all 29 levels of the 4f$^3$($^4$I)6p, and all 8 levels 4f$^3$($^4$I)6s sub-configurations. Despite the 4f$^4$~$^5$I - 4f$^3$($^4$I)5d transitions lying within the line-rich and blend-rich visible regions, their classifications were possible due to their high relative intensities and the large number of 4f$^3$($^4$I)5d~-~4f$^3$($^4$I)6p transitions observed and straightforwardly classified in the line-sparse and generally blend-free UV region.

\subsubsection{\texorpdfstring{The 4f$^3$($^5$G)5d levels and Cowan code calculations}{The 4f3(5G)5d levels and Cowan code calculations}}
A group of lines with the highest SNRs in spectrum E around 35500~cm$^{-1}$ showed the 4f$^4$ - 4f$^3$5d isotope profile and resembled relative intensities of the 4f$^4$~$^5$I - 4f$^3$($^4$G)5d~$^5$H transitions predicted from \cite{gaigalas2019extended}. All of the 4f$^3$($^4$G)5d~$^5$H levels were found around 2500~cm$^{-1}$ below the energies calculated in \cite{gaigalas2019extended}, this offset was then expected for the other 4f$^3$($^4$G)5d levels, which led to the classification of weaker 4f$^4$~$^5$I - 4f$^3$($^4$G)5d transitions from a few levels of the 4f$^3$($^4$G)5d~$^5$G and 4f$^3$($^4$G)5d~$^5$I terms. 

At this stage, many observable levels predicted by \cite{gaigalas2019extended} appeared highly mixed, which was indicated by their missing or multiply assigned \textit{LS}-coupling labels. Parameterised Cowan code calculations were then performed including the new levels established for the 4f$^3$($^4$I)5d, 4f$^3$($^4$I)6s, 4f$^3$($^4$I)6p, and 4f$^3$($^4$G)5d sub-configurations. \textit{LS} labels of a handful of 4f$^3$5d levels were re-assigned using the newly calculated eigenvector compositions, and the improved transition probabilities enabled the identification of more 4f$^3$5d levels, all of these were assigned to the 4f$^3$($^4$F)5d, 4f$^3$($^2$H2)5d, 4f$^3$($^4$G)5d, and 4f$^3$($^2$K)5d sub-configurations. Remaining transitions of the experimentally unknown 4f$^3$5d levels to the ground term 4f$^4$~$^5$I were expected to be very weak in the PDL spectrum, so the focus shifted to identifying other low-lying levels of the 4f$^4$ configuration. 

\subsubsection{\texorpdfstring{The 4f$^4$ levels}{4f4 levels}}
Levels of the low-lying $^5$F, $^5$G, $^5$H4, and $^3$K2 terms of the 4f$^4$ configuration were identified. Their transitions from the 4f$^3$5d levels were generally much weaker and lay within the most line-rich spectral regions in the visible. Isotope profiles and relative intensities of these transitions were much more uncertain due to low SNRs and high chances of significant line blending. Cowan code calculations were also less well constrained for the 4f$^4$ configuration due to the lack of identified levels of this configuration. This was the final and most lengthy stage of the analysis, where Nd VS and stellar grating spectra became vital.

\begin{figure}
    \includegraphics[width=\linewidth]{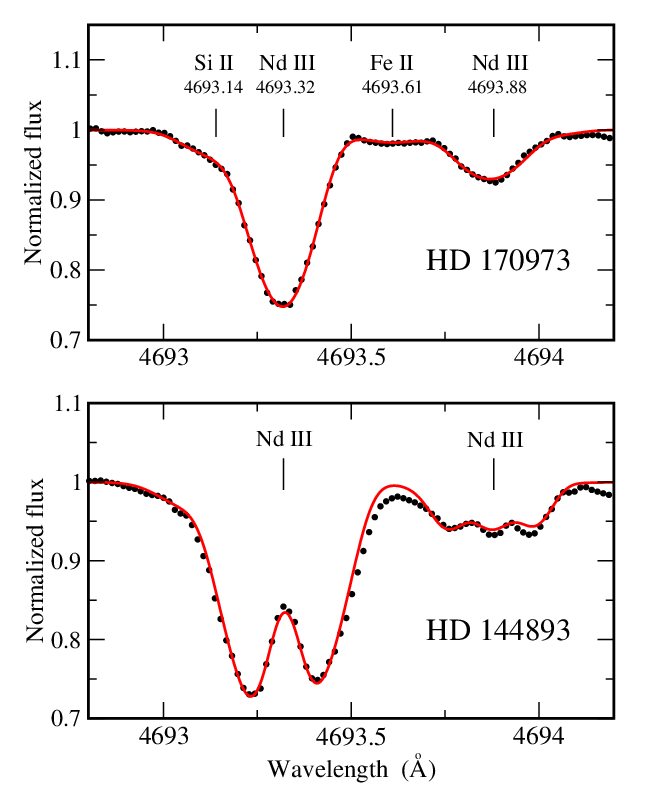}
    \caption{Observed (black dotted line) and synthesised (red line) spectra between 4692.5 - 4694.5~\AA{} of HD~170973 and HD~144893 with surface magnetic fields \bz$<$1~kG and \bz=8.8~kG respectively, containing two newly classified transitions of Nd~III: 4f$^4$~$^5$F$_5$ - 4f$^3$($^4$F)5d~$^5$F$_5$ at 4693.32~\AA{} and 4f$^4$~$^5$G$_5$ - 4f$^3$($^4$G)5d~$^5$G$_6$ at 4693.88~\AA{}.} \label{fig5}
\end{figure}

Initially, progress was made identifying 4f$^4$~$^5$G$_5$, where the 5 lines observed in the Nd-Ar PDL spectrum (4 of which were at SNRs less than 10, 2 of which were blended) were supported by the spectral syntheses of HD~170973 and HD~144893 using transition probabilities and Land\'{e} $g$-factors predicted by the Cowan code. A nearly blend-free example is shown in Figure \ref{fig5} containing the 4f$^4$~$^5$G$_5$ - 4f$^3$($^4$G)5d~$^5$G$_6$ transition (SNR 5 in the PDL spectrum) at 4693.88~\AA{}, where corroboration with stellar spectra is also shown for the later classified transition of 4f$^4$~$^5$F$_5$ - 4f$^3$($^4$F)5d~$^5$F$_5$ at 4693.32~\AA{}. 
The average Nd abundance from 4 mostly blend-free 4f$^4$~$^5$G$_5$ lines in the spectrum of HD~170973 was estimated at $\log\varepsilon_{\rm Nd}$=5.65$\pm$0.09, in agreement with the reference value 5.63$\pm$0.14 (see Section~\ref{sec:stellar spectra}). Except for the 4693.87~\AA{} line, all lines of 4f$^4$~$^5$G$_5$ in the spectrum of HD~144897 were blended by Zeeman components of nearby lines, but 3 of them provided an abundance estimate at $\log\varepsilon_{\rm Nd}$=5.66$\pm$0.13, which again was in agreement with average Nd abundance 5.59$\pm$0.20 derived in this star by \cite{ryabchikova2006rare} from other Nd III lines. Such abundance estimates greatly supported the identification of 4f$^4$~$^5$G$_5$ and other 4f$^4$ levels.

The classification of transitions to 4f$^4$~$^5$G$_5$ provided a useful guide on the expected relative intensities and SNRs of transitions from 4f$^3$5d to other nearby low-lying 4f$^4$ levels observed in the Nd-Ar PDL spectra. Subsequently, the 4f$^4$~$^3$K$_{6,7}$2 and 4f$^4$~$^5$F$_5$ levels\footnote{The trailing numbers of a level label are used to distinguish recurrent terms of equivalent electrons, e.g., 4f$^4$~$^3$K$_{6,7}$2 refers to the $J=6,7$ levels of the second $^3$K term of the 4f$^4$ configuration.} were confidently identified with support from H. M. Crosswhite's unclassified list of 643 Nd~III lines. Most notably, these lines from her sliding spark grating spectra indicated several transitions that were weak, blended, and/or not fitted in the Nd-Ar PDL spectra. The 4 levels belonging to 3 separate terms of the 4f$^4$ configuration improved our Cowan code parameterisation, and the identifications of 4f$^4$~$^3$K$_{8}$2, 4f$^4$~$^5$G$_{3,4}$, and 4f$^4$~$^3$H$_5$4 followed with confirmations from stellar spectra, which also enabled the identification of more levels of the 4f$^3$($^4$F)5d, 4f$^3$($^2$H2)5d, 4f$^3$($^4$G)5d and 4f$^3$($^4$D)5d (assigned) sub-configurations.

Identification of the 4f$^4$~$^5$F$_{2,3,4}$ levels proved more challenging. The transitions 4f$^4$~$^5$F$_J$~-~4f$^3$($^4$S)5d~$^5$D$_{J-1}$ ($J>1$) were tentatively classified prior to the establishment of their energy levels, using their high intensities, isotope profiles, agreement in intensity and wavenumber with calculations, and presence in H. M. Crosswhite's unclassified Nd~III line list and the stars. But the 4f$^4$~$^5$F$_J$~-~4f$^3$($^4$S)5d~$^5$D$_{J}$ transitions were expected around the noise level with too many candidate classifications, so the levels of 4f$^4$~$^5$F and 4f$^3$($^4$S)5d~$^5$D terms could not be established using only the transitions between them. In looking for other evidence to resolve these ambiguities, the 4f$^3$($^4$F)6p~$^5$G$_{2,3,4,5}$ levels proved useful, which were identified using the VS grating spectra with the strongest lines to 4f$^3$($^4$F)5d~$^5$H$_{3,4,5,6}$ also observed in the PDL spectrum respectively (the same applied in the identifications of the 4f$^3$($^2$H2)6p levels). The 4f$^3$($^4$F)6p~$^5$G levels aided the identification of the 
4f$^3$($^4$F)5d~$^5$F$_{2,3,4}$ levels, which had transitions to 4f$^4$~$^5$F$_{2,3,4}$ at intermediate SNRs. Even when additionally considering transitions from 4f$^3$($^4$F)5d~$^5$F$_{2,3,4}$, several attempts were made at classifying the 4f$^4$~$^5$F$_{2,3,4}$ - 4f$^3$5d transitions in the Nd-Ar PDL spectra before an agreement could be reached with lines observed in the stellar spectra.

\section{Outlook}
This extended analysis of the four low-lying configurations 4f$^4$, 4f$^3$5d, 4f$^3$6s, and 4f$^3$6p of Nd~III will enable laboratory measurements of their lifetime and branching fractions for accurate transition probabilities. The experimentally established energy levels will improve future atomic structure calculations of Nd~III by serving as benchmarks and semi-empirical constraints. 

Attempts to identify levels of the doubly-excited configurations of Nd~III were unsuccessful, the strongest predicted transitions originating from 4f$^2$5d$^2$ are to the still unknown levels of the 4f$^3$5d configuration. The remaining unknown 4f$^4$~-~4f$^3$5d transitions are expected to be very weak in the recorded Nd-Ar PDL FT spectra, with too many ambiguities in the expected intensity patterns and wavenumber ranges. Further large-scale experimental investigations of Nd~III atomic structure would require extensions of the presented experimental and analysis methods, e.g., altering Nd discharge conditions to better populate higher-lying Nd~III energy levels and to separate Nd~III lines from other species, and/or establishing isolated energy level systems with strong lines which would eventually be connected to the current network of levels to the ground level.

Energy levels of the 4f$^3$($^4$I)6d and 4f$^3$($^4$I)7s sub-configurations have been identified in the Nd-Ar PDL FT spectra together with aid from the Nd VS grating spectra, they lie about 6000~cm$^{-1}$ lower than predicted by \cite{gaigalas2019extended}. Levels of the 4f$^3$5f configuration are also being established from the classification of the 4f$^3$5d - 4f$^3$5f transitions in another set of Nd grating spectra recorded in the range 820-1159~\AA{}. These results, totaling more than 100 energy levels, are being prepared for publication, where the energy level and transition probability calculations by Cowan's codes will also be included in full.

Many energy levels of Nd~I-III remain unknown experimentally, as more than $10^4$ lines are still unclassified in the Nd-Ar PDL spectra. The Penning discharge lamp design has shown to be a suitable source of Nd spectra, with particular enhancements of Nd~II and Nd~III populations compared to the alternative HCL at running conditions which maximised the Nd~III resonance line intensities. For further investigations of Nd and other lanthanide elements, high-resolution spectroscopy of varying plasma conditions, accurate semi-empirical atomic structure calculations, and rigorous line list extraction methods are emphasised to be key strategies.

\section{Summary}
144 energy levels of Nd~III have been determined from the classification of 433 Nd~III transitions measured by Fourier transform spectroscopy of Nd Penning lamp discharges between 11500-54000~cm$^{-1}$ (8695-1852 \AA) supplemented by grating spectroscopy of Nd vacuum sliding sparks and Nd-rich stars. 39 previously published energy levels were confirmed and 105 new levels of the 4f$^4$, 4f$^3$5d, 4f$^3$6s, and 4f$^3$6p configurations are reported here for the first time. These results are the most extensive and most accurate (to a few 10$^{-3}$~cm$^{-1}$) Nd~III energy level and transition wavenumber data to date, which will enable the wider and more reliable application of Nd~III atomic data in astronomy.
\begin{acknowledgments}
    This work was supported by the STFC of the UK and the research project FFUU-2022-0005 of the Institute of Spectroscopy of the Russian Academy of Sciences. The authors are grateful to J.-F. Wyart and Prof. W.-\"{U}. Tchang-Brillet for providing the Nd vacuum sliding spark grating plates recorded at NIST, and to Prof. C. R. Cowley for sharing the unpublished Nd~III line lists of H. M. Crosswhite.
\end{acknowledgments}

\bibliography{bibliography}{}
\bibliographystyle{aasjournal}




\end{document}